\documentclass[cup6b]{cuptest}
\usepackage{bm}
\usepackage{natbib}
\usepackage{graphicx}

\def\@biblabel#1{\relax}

\newcommand{\eqnsize}{}

\newcommand{\myequation}[1]{ \begin{eqnarray} {{#1}}   \end{eqnarray} }

\newcommand{\vc}[1]{{\bf #1}}
\newcommand{\bn}{\hat{\bf n}}
\newcommand{\bx}{{\bf x}}
\newcommand{\bk}{{\bf k}}
\newcommand{\bl}{{\bf l}}
\newcommand{\sig}[1]{{\boldsymbol \sigma}_{#1}}
\newcommand{\mat}[4]{\left( \begin{array}{cc} 
	#1 \! & #2 \\
	#3 \! & #4 \end{array} \right)}

\title[]
      {Lecture Notes on CMB Theory:\\ From Nucleosynthesis to Recombination}
\author{Wayne Hu}
\date{\today}

\begin{document}

\pagenumbering{roman}
\maketitle
\setcounter{tocdepth}{2}
\tableofcontents
\cleardoublepage
\pagenumbering{arabic}

\chapter{CMB Theory from Nucleosynthesis to Recombination}
\section{Introduction}

These lecture notes comprise an introduction to the well-established physics and
phenomenology of the cosmic microwave background (CMB) between big bang nucleosynthesis
and recombination.    We take our study through recombination since most of the temperature and polarization anisotropy observed in
the CMB formed during the recombination epoch when free electrons became bound into
hydrogen and helium.

The other reason for considering only this restricted range from nucleosynthesis to recombination is that these notes are meant to complement the other  lectures  
of  the XIX Canary Island Winter School of Astrophysics.  While they are self-contained
and complete in and of themselves, they omit important topics that will be covered elsewhere in this volume:
namely inflation (Sabino Matarrese), observations (Bruce
Partridge), statistical analysis (Licia Verde),
 secondary anisotropy (Matthias Bartelmann), and non-Gaussianity
(Enrique Martinez-Gonzalez).   

Furthermore the approach taken here of introducing
only as much detail as necessary to build physical intuition is more suitable as a general
overview rather than a rigorous treatment for training specialists.    As such these notes complement the more formal lecture notes for the Trieste school which may anachronistically be viewed as a continuation of these notes
in the same notation (\citealt{Hu04a}).  

The outline of these notes are as follows.
We begin in \S\ref{sec:thermal} with a brief thermal history of the CMB.  We discuss the
temperature and polarization 
anisotropy and acoustic peaks from recombination in \S\ref{sec:anisotropy}-\ref{sec:polarization}
and conclude in \S\ref{sec:discussion}.
We take units throughout with $\hbar=c=k_{B}=1$ and illustrate effects in the
standard cosmological constant cold dark matter universe with adiabatic inflationary
initial conditions ($\Lambda$CDM).

\begin{figure}[tb]
\begin{center}
\includegraphics[width=4in]{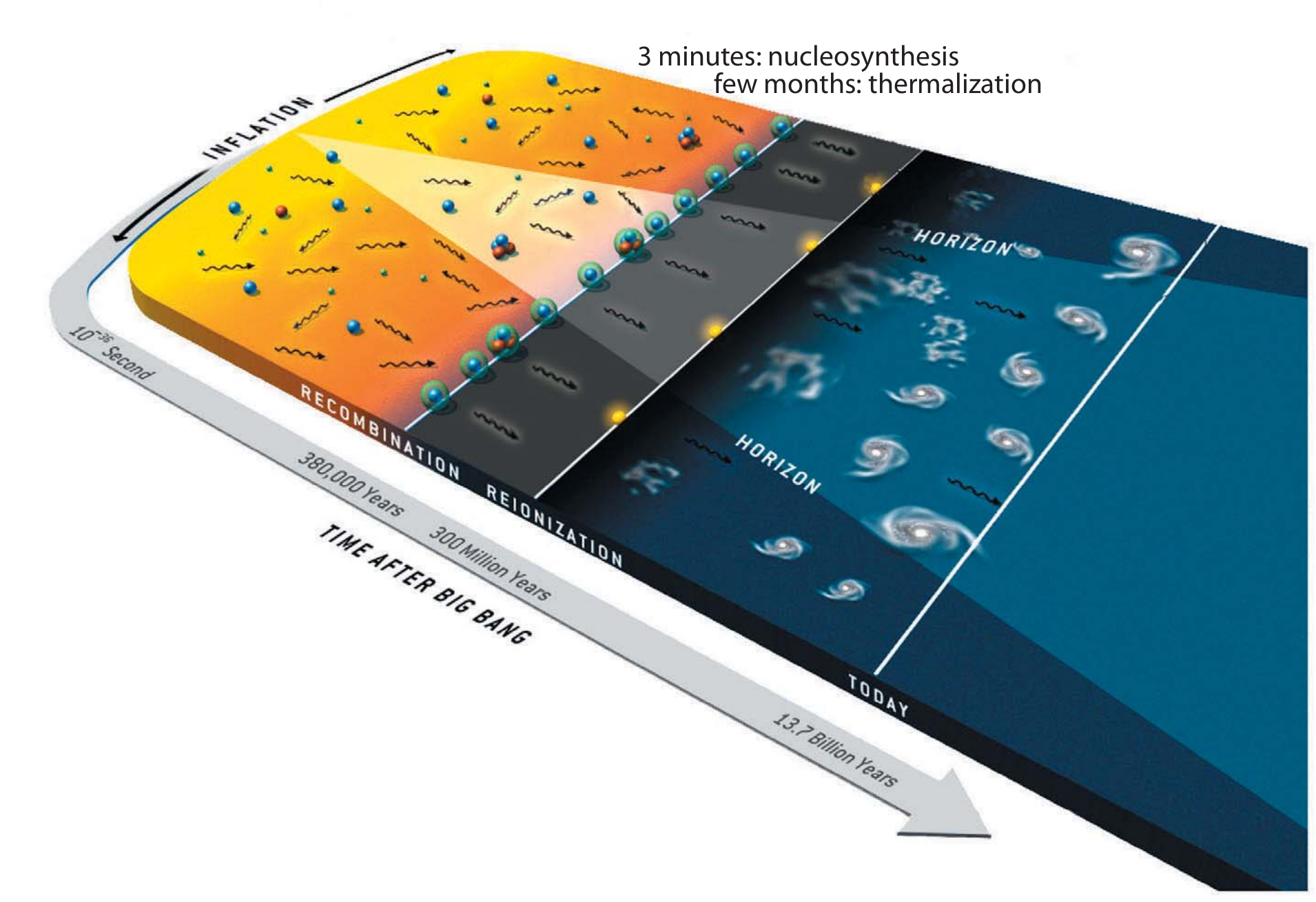}
\caption{A brief thermal history: nucleosynthesis, thermalization, recombination and reionization. Adapted from \cite{HuWhiSciAm}.} \label{fig:timeline}
\end{center}
\end{figure}
\section{Brief Thermal History}
\label{sec:thermal}

In this section, we discuss the major events in the thermal history of the CMB.  
We begin in \S\ref{sec:bbn} with the formation of the light elements and the original prediction
of relic radiation.   We continue in \S\ref{sec:thermalization} with the processes that thermalize
the CMB into a blackbody.   Finally in \S\ref{sec:recombination}, we discuss the recombination epoch where the
main sources of temperature and polarization anisotropy lie.

\subsection{Nucleosynthesis and Prediction of the CMB}
\label{sec:bbn}

Let us begin our  brief thermal history with the relationship between the
CMB and the abundance of light elements established at an energy scale of $10^2$ keV, time
scale of a few minutes, temperature of $10^9$K 
and redshift of $z \sim 10^8-10^9$.  This is the epoch of nucleosynthesis, the formation
of the light elements.  
The qualitative features of nucleosynthesis are set by the low baryon-photon number density of our universe.  Historically, the sensitivity to this ratio was used by 
Gamow  and collaborators in the late 1940's to predict the existence and estimate
the temperature of the CMB.   Its modern use is the opposite: with the photon
density well measured from the CMB spectrum, the abundance of light elements
determines the baryon density.

 At the high temperature and densities of nucleosynthesis,
radiation is rapidly thermalized to a perfect black body and the photon number
density is a fixed function of the temperature $n_\gamma \propto T^3$ (see below).  
Apart from epochs in which energy from particle annihilation or other processes
is dumped into the radiation, the baryon-photon number density ratio remains constant.

Likewise nuclear statistical equilibrium, while satisfied, makes the abundance of the light
elements of mass number $A$ follow the expectations of a Maxwell-Boltzmann distribution
for the phase space occupation number
\begin{equation}
 f _A= e^{-(m_A-\mu_A)/T } e^{-p_A^{2}/2m_AT } \,,
 \end{equation}
where $p_A$ is the particle momentum, $m_{A}$ is the rest mass, and  $\mu_A$ is
the chemical potential.  Namely their number density
\begin{eqnarray}
n_A& \equiv &  g_A \int  {d^3 p_A \over (2\pi)^3} f_A \nonumber\\
&
= & g_A ({m_A T \over 2\pi} )^{3/2} e^{(\mu_A - m_A)/T}  \,.
\label{eqn:maxwellboltzmann}
\end{eqnarray}
Here  $g_A$ is the degeneracy factor.   In equilibrium, the chemical potentials of the various elements
are related to those of the proton and neutron by
\begin{equation}
\mu_A = Z \mu_p + (A-Z)\mu_n \,,
\end{equation}
where $Z$ is the charge or number of protons.   Using this relation, the abundance
fraction
\begin{equation}
X_A  \equiv A {n_A \over n_b } = A^{5/2} g_A 2^{-A} \left[ \left( { 2\pi T \over m_b} \right)^{3/2}{2 \zeta(3) \eta_{b\gamma} \over \pi^2} 
   \right]^{A-1}    e^{B_A/T} 
   X_p^Z X_n^{A-Z} \,,
   \label{eqn:nse}
   \end{equation}
   where $X_p$ and $X_n$ are the proton and neutron abundance, $\zeta(3)\approx 1.202$,
   and $n_b$ is the baryon number density.  The two controlling quantities are
   the binding energy 
   \begin{equation}
   B_A = Z m_p + (A-Z) m_n - m_A \,,
   \end{equation}
   and the baryon-photon number density ratio
$\eta_{b\gamma}=n_{b}/n_{\gamma}$.  That the latter number is of order $10^{-9}$ in our universe means
that light elements form only well after the temperature has dropped below the binding
energy of each species.   Nuclear statistical equilibrium holds until the reaction rates drop below the
expansion rate.  At this point, the abundance freezes out and remains constant. 

Gamow's back of the envelope estimate (\citealt{Gam48}, refined by 
 \citealt{AlpHer48}) was to consider the neutron capture reaction that
forms deuterium
\begin{equation}
n + p \leftrightarrow d + \gamma
\end{equation}
with a binding energy of $B_2 = 2.2$MeV.   Given Eqn.~(\ref{eqn:nse}) 
\begin{equation}
X_2 = {3 \over \pi^2}  \left( { 4\pi T \over m_b} \right)^{3/2}  \eta_{b\gamma}
  \zeta(3) e^{B_2/T} 
   X_p X_n\,,
\end{equation}
a low baryon-photon ratio, and $X_n \approx X_p \approx 1/2$ for estimation purposes, the critical temperature for
deuterium formation is  $T \sim 10^9$K.  In other words the low baryon-photon 
ratio means that there are sufficient numbers of photons to dissociate deuterium until
well below $T \sim B_2$.  
Note that this condition is
only logarithmically sensitive to the exact value of the baryon-photon ratio chosen
for the estimate and so the reasoning is not circular. 

Furthermore, that we observe deuterium at all and not all helium and heavier
elements means that the reaction must have
frozen out at near this temperature.  Given the thermally averaged cross section of 
\begin{equation}
\langle \sigma v \rangle \approx 4.6 \times 10^{-20} {\rm cm}^3 {\rm s}^{-1}\,,
\end{equation} 
 the freezeout condition
\begin{equation}
n_b \langle \sigma v \rangle \approx H \approx t^{-1}\,,
\end{equation}
and the time-temperature relation $t(T=10^{9}{\rm K}) \approx 180$s from the 
radiation dominated Friedmann equation, we obtain an estimate of the baryon 
number density 
\begin{equation}
n_b( T=10^9{\rm K}) \sim 1.2 \times 10^{17} {\rm cm}^{-3} \,.
\end{equation}
Comparing this density to the an observed current baryon density and requiring that
$n_b \propto a^{-3}$ and $T \propto a^{-1}$ yields the current temperature of the
thermal background.  For example, taking the modern value of $\Omega_b h^2 \approx 0.02$
and $n_b(a=1)=2.2 \times 10^{-7}$cm$^{-3}$ yields the rough estimate
\begin{equation}
T(a=1) \approx 12{\rm K} \,.
\end{equation}
This value is of the same order of magnitude as the observed CMB temperature $2.725$K 
as well as the original estimates
(\citealt{AlpHer48,DicPeeRolWil65}).   Modern day estimates of the baryon-photon ratio
also rely on deuterium (e.g. \citealt{Tytetal00}).  
We shall see that the CMB has
its own internal measure of this ratio from the acoustic peaks.   Agreement between the
nucleosynthesis and acoustic peak measurements argue that the baryon-photon ratio
has not changed appreciably since $z \sim 10^{8}$.

\begin{figure}[tb]
\begin{center}
\includegraphics[width=4in]{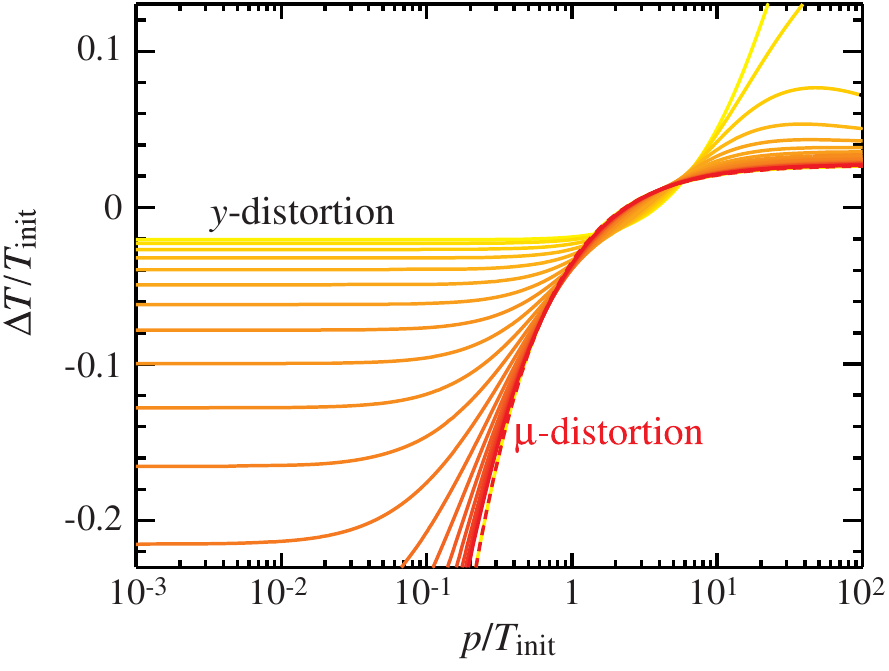}
\caption{Comptonization process.  Energy injected into the CMB through heating of the
electrons is thermalized by Compton scattering.   Under the Kompaneets equation, a $y$-distortion first forms as low frequency photons gain energy from the electrons.  After multiple
scatterings the distribution is thermalized to a chemical potential or $\mu$-distortion.
 Adapted from \cite{HuThesis}.} \label{fig:comptonization}
\end{center}
\end{figure}

\subsection{Thermalization and Spectral Distortions}
\label{sec:thermalization}

Between nucleosynthesis and recombination, processes that create and destroy photons
and hence thermalize the CMB fall out of equilibrium.    The lack of spectral distortions
in the CMB thus constrains any process that injects energy or photons into the plasma
after this epoch.   

In a low baryon-photon ratio universe, the main thermalization process is double, also
known as radiative, Compton scattering
\begin{equation}
e^- + \gamma \leftrightarrow e^- + \gamma +{\gamma} \,.
\end{equation}
The radiative Compton scattering rate becomes insufficient to maintain a blackbody
at a redshift of (\citealt{DanDes82})
\begin{equation}
z_{\rm therm}= 2.0 \times 10^{6}(1-Y_{p}/2)^{-2/5}\left( { \Omega_{b}h^{2} \over 0.02} \right)^{-2/5}
\end{equation}
corresponding to a time scale of order a few months.

After this redshift, energy or photon injection appears as a spectral distortion in 
the spectrum of the CMB.    The form of the distortion is determined by Compton scattering
\myequation{ e^- + \gamma \leftrightarrow e^- + \gamma\,,}
since it is still sufficiently rapid compared with respect to the expansion
while hydrogen remains ionized.
  Because a blackbody has a definite number density of photons
at a given temperature, energy exchange via Compton scattering alone can only produce a 
Bose-Einstein spectrum for the photon distribution function 
\begin{equation}
f = {1 \over  e^{(E-\mu)/T} - 1 }  
\end{equation}
with $\mu$ determined by the conserved number density and temperature.
The evolution to a Bose-Einstein distribution  is determined by solving the Kompaneets equation (\citealt{ZelSun69}).  
The Kompaneets equation is Boltzmann equation for
Compton scattering in a homogeneous medium.  It includes the effects of
electron recoil and the second order Doppler shift that exchange energy between the
photons and electrons.   If energy is injected into the plasma to heat the
electrons, Comptonization will try to redistribute the energy to the CMB.  
Consequently the first step in Comptonization is a so called ``$y$ distortion'' in 
the spectrum  as
low energy photons in the Rayleigh Jeans tail of the CMB gain energy from the
second order Doppler shift (see Fig.~\ref{fig:comptonization}).

After multiple scatterings, the spectrum settles into the Bose-Einstein 
form with a ``$\mu$ distortion.'' 
The transition occurs when the energy-transfer weighted optical depth
approaches unity, \emph{i.e.}  $\tau_K(z_K)=4 \int dt  n_e \sigma_T T_e/m_e=1$ where
$n_e$ is the electron density $\sigma_T$ is the Thomson cross section, $T_e$ is
the electron temperature and $m_e$ is the electron mass 
\begin{equation}
z_K \approx 5.1 \times 10^4 ( 1- Y_p/2)^{-1/2} \left({\Omega_b h^2 \over 0.02}\right)^{-1/2}\,.
\end{equation}
Above this redshift energy injection creates a $\mu$ distortion, below a $y$ distortion.

\begin{figure}[tb]
\begin{center}
\includegraphics[width=4in]{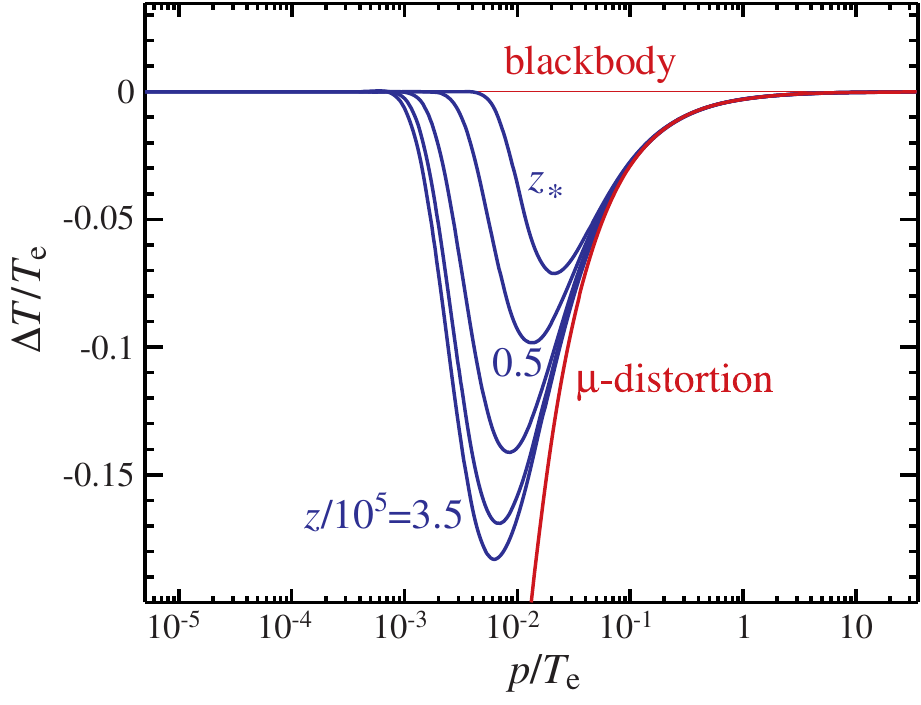}
\caption{Low frequency thermalization by bremsstrahlung. Creation and absorption of
photons  at low frequencies by bremsstrahlung bring a chemical potential distortion back to blackbody at the
electron temperature $T_{e}$
 between $z_{K}$ and recombination.  
Adapted from \cite{HuSil93}.} \label{fig:thermalization}
\end{center}
\end{figure}

At very low frequencies, bremsstrahlung
\begin{equation}
e^- + p  \leftrightarrow e^- + p +{\gamma}
\end{equation}
 is still efficient in creating and absorbing
photons.  In Fig.~\ref{fig:thermalization} we show the how a $\mu$-distortion 
continues to evolve until recombination 
 which brings the low frequency spectrum back to a blackbody but now at the electron
 temperature.  

\begin{figure}[tb]
\begin{center}
\includegraphics[width=4in]{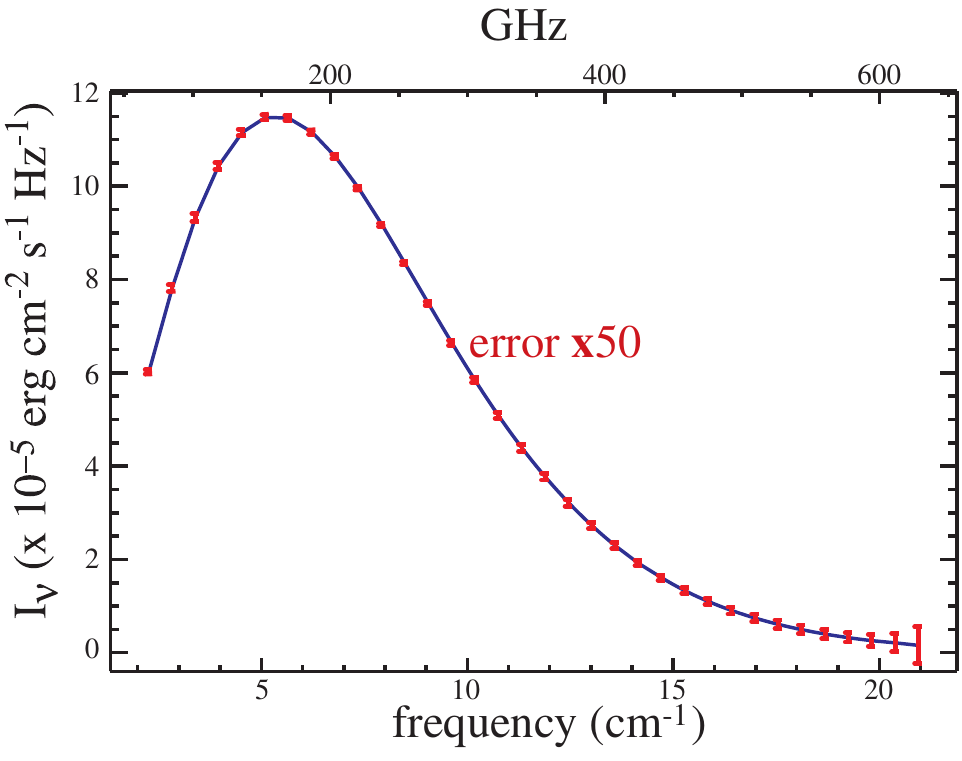}
\caption{CMB spectrum from FIRAS.  No spectral distortions from a blackbody have
been discovered to date.} \label{fig:firas}
\end{center}
\end{figure}

The best limits to date are from COBE FIRAS from intermediate to high
frequencies: $|\mu| < 9 \times 10^{-5}$ and $|y|<1.5 \times 10^{-5}$ at 95\% confidence
(see Fig~\ref{fig:firas} and \citealt{Fixetal96}).  After subtracting out galactic  emission,
no spectral distortions of any kind are detected and the spectrum appears to be a
perfect blackbody of $\bar T=2.725\pm 0.002$K (\citealt{Matetal99}).

\subsection{Recombination}
\label{sec:recombination}

While the recombination process 
\begin{equation}
{p} + { e^{-} }\leftrightarrow {H} + \gamma \,,
\end{equation}
is rapid compared to the expansion, 
the ionization fraction obeys an equilibrium distribution just like that considered
for light elements for nucleosynthesis or the CMB spectrum thermalization.  As in the
former two processes, the qualitative behavior of recombination is determined
by the low baryon-photon ratio of the universe.

Taking number densities of the Maxwell-Boltzmann form of Eqn.~(\ref{eqn:maxwellboltzmann}),
we obtain
\myequation{ {n_{p} n_{e} \over n_{H}} \approx e^{-{B}/T} \left( {m_{e} T \over 2\pi} \right)^{3/2}
	e^{({\mu_{p}} +{\mu_{e}}-{\mu_{H}})/T }\,,
	\label{eqn:ionizationdirect}}
{where ${B} = m_{p}+m_{e}-m_{H}= 13.6$eV is the {binding energy} and we have
set
$g_p=g_e={1\over 2} g_H=2$.  Given the vanishingly small chemical potential of the
photons, $\mu_{p}+\mu_{e}=\mu_{H}$ in equilibrium.}

\begin{figure}[tb]
\begin{center}
\includegraphics[width=4in]{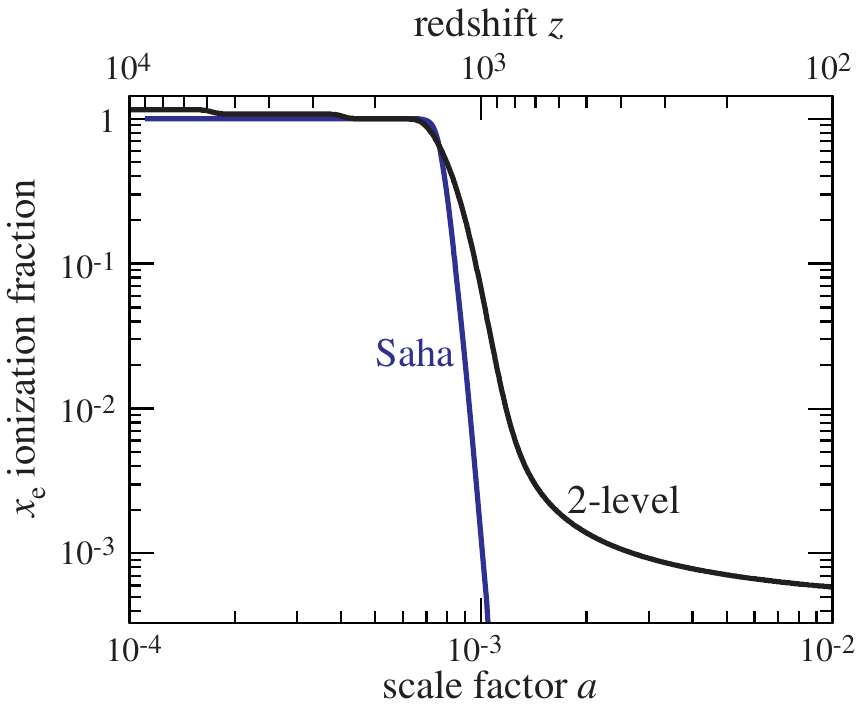}
\caption{Hydrogen recombination in Saha equilibrium vs. the calibrated 2-level calculation of RECFAST.
The features before hydrogen recombination are due to helium recombination.} \label{fig:recombination}
\end{center}
\end{figure}

{Next, defining the ionization fraction for a hydrogen only plasma}
\begin{eqnarray}
n_{p} &=& n_{e}  = x_{e} n_{b}\,, \nonumber\\
n_{H} & =& n_{b} - n_{p} = (1-x_{e})n_{b}\,,
\end{eqnarray}
we can rewrite Eqn.~(\ref{eqn:ionizationdirect}) as the Saha equation
\begin{eqnarray}
{n_e n_p \over n_H n_b} & = & {{x_e^2} \over 1-{x_e}}  =\frac{1}{n_b} \left( { m_e {T} \over 2\pi} \right)^{3/2} e^{-{B}/{T} }\,. 
\end{eqnarray}

Recombination occurs at a temperature substantially lower than $T=B$ again because
of the low baryon-photon ratio of the universe.   We can see this by rewriting the Saha equation
in terms of the photon number density
\begin{eqnarray}
{x_{e}^{2} \over 1-x_{e}}
= e^{-B/kT} {\pi^{1/2}\over 2^{5/2}\eta_{b\gamma} \zeta(3)}
\left( { m_{e}c^{2}\over k T } \right)^{3/2} \,.
\end{eqnarray}
Because $\eta_{b\gamma}\sim 10^{-9}$, the Saha equation implies that
the medium only becomes substantially neutral at a temperature of $T \approx 0.3$eV
or at a redshift of $z_{*} \sim 10^{3}$.   At this point, there are not enough photons
in the even in the Wien tail above the binding energy to ionize hydrogen.
We plot the Saha solution in Fig.~\ref{fig:recombination}.

Near the epoch of recombination, the recombination rates become insufficient to maintain
ionization equilibrium.   There is also a small contribution from helium recombination 
that must be added.
The current standard for following the non-equilibrium
 ionization history is RECFAST (\citealt{SeaSasSco99})
which employs the traditional two-level atom calculation of \cite{Pee68} but
alters the hydrogen case $B$ recombination rate $\alpha_{B}$ to fit the results of 
a multilevel atom.   More specifically, RECFAST solves a coupled system of
equations for the ionization fraction $x_i$ in singly ionized hydrogen and helium ($i=$ H, He)
\begin{equation}
{d x_i \over d\ln a} = {\alpha_B C_i n_{Hp} \over H} \left[ s(x_{\rm max}-x_i) - x_i x_e \right]\,,
\end{equation}
where $n_{Hp} = (1-Y_p)n_b$ is the total hydrogen plus proton number density accounting for the
helium mass fraction $Y_p$,
$x_{e} \equiv n_e/n_{Hp} =\sum x_i$ is the total ionization fraction,  $n_e$ is the 
free electron density, $x_{\rm max}$ is the maximum $x_i$ achieved through full
ionization, 
\begin{eqnarray}
s & = & {\beta \over n_{Hp}} e^{-B_{1s} /  T_b}\,,\nonumber\\
C_i^{-1} & = &  1 + {\beta \alpha_B e^{-B_{2s}/ T_b} \over \Lambda_\alpha + \Lambda_{2s1s} }\,,\nonumber\\
\beta & = & g_{\rm rat} \left( {  T_b m_e \over 2\pi  } \right)^{3/2} \,,
\end{eqnarray}
with $g_{\rm rat}$ the ratio of statistical weights,
$T_b$  the baryon temperature, $B_{L}$ the binding energy of the $L$th level,
$\Lambda_\alpha$ the rate of redshifting out of the Lyman-$\alpha$ line 
corrected for the energy difference
between the $2s$ and $2p$ states
\begin{equation}
\Lambda_\alpha = {1 \over \pi^2} \left( { B_{1s}-B_{2p}  }\right)^3 e^{-(B_{2s}-B_{2p})/ T_b} 
{H \over (x_{\rm max} - x_i) n_{Hp}} 
\end{equation}
and $\Lambda_{2s1s}$ as the rate for the 2 photon $2s-1s$ transition. For reference, for hydrogen $B_{1s} = 13.598$eV, $B_{2s}=B_{2p}=B_{1s}/4$,
$\Lambda_{2s1s}=8.22458 s^{-1}$, $g_{\rm rat}=1$, $x_{\rm max}=1$.  For helium
$B_{1s}= 24.583$eV, $B_{2s}=3.967$eV, $B_{2p}=3.366$eV,  $\Lambda_{2s1s}=51.3 s^{-1}$,
$g_{\rm rat}=4$, $x_{\rm max}=Y_p/[4(1-Y_p)]$. 

Note that 
if the recombination rate is faster than the expansion rate $\alpha_B C_i n_{Hp}/ H \gg 1$, the ionization solutions for $x_i$ reach the Saha 
equilibrium $d x_{i}/d\ln a=0$.  In this case $s( x_{\rm max}-x_i) = x_i x_e$ or
\begin{eqnarray}
x_i &=& {1 \over 2} \left[ \sqrt{ (x_{ei}+s)^2 + 4 s x_{\rm max}} - ({x_{ei} + s}) \right] \\
   &=& x_{\rm max} \left[  1 - {x_{ei} + x_{\rm max}  \over s} \left(  1 - { x_{ei} + 2 x_{\rm max} \over s } \right) 
   + ...\right]\,, \nonumber
\end{eqnarray}
where $x_{ei}= x_e - x_i$ is the ionization fraction excluding  the species. The recombination of hydrogenic doubly ionized helium is here handled purely through the
Saha equation with a binding energy of ${1/4}$ the $B_{1s}$ of hydrogen and
$x_{\rm max}=Y_p/[4(1-Y_p)]$.  The case $B$ recombination coefficients as a function of $T_b$
are given in  \cite{SeaSasSco99} as is the strong thermal coupling between $T_b$ and
$T_{\rm CMB}$.  The multilevel-atom fudge that RECFAST introduces
is to replace the hydrogen $\alpha_B \rightarrow 1.14 \alpha_B$ independently of cosmology.
While this fudge suffices for current observations, which approach the $\sim1\%$ level, 
the recombination standard
will require improvement if CMB anisotropy predictions are to reach an accuracy of $0.1\%$
(e.g.~\citealt{SwiHir07,WonMosSco07}).


The phenomenology of CMB temperature and polarization anisotropy is primarily governed by the redshift of recombination
$z_* = a_*^{-1}-1$ when most of the contributions originate.   This redshift though
carries little dependence on standard cosmological parameters.
  This insensitivity
follows from the fact that recombination proceeds rapidly once  $B_{1s}/T_b$ has reached
a certain threshold as the Saha equation illustrates.   
Defining the redshift of recombination as the epoch 
 at
which the Thomson optical depth during recombination (\emph{i.e.}~excluding reionization) reaches unity,
$\tau_{\rm rec}(a_{*})=1$, a fit to the recombination calculation
 gives (\citealt{Hu04b})
\begin{equation}
a_*^{-1}  \approx 1089 \left({\Omega_m h^2 \over  0.14} \right)^{0.0105} 
				\left( {\Omega_b h^2 \over 0.024} \right)^{-0.028} \,,
\label{eqn:zstarapp}
\end{equation}
around a fiducial model
of $\Omega_m h^2=0.14$ and $\Omega_bh^2=0.024$.

The universe is known to be reionized at low redshifts due to the lack of a Gunn-Peterson
trough in quasar absorption spectra.  Moreover, large angle CMB polarization detections
(see Fig.~\ref{fig:spectra}) suggest that this transition back to full ionization occurred around
$z\sim10$ leaving an extended neutral period between recombination and reionization.

\begin{figure}[tb]
\begin{center}
\includegraphics[width=4.5in]{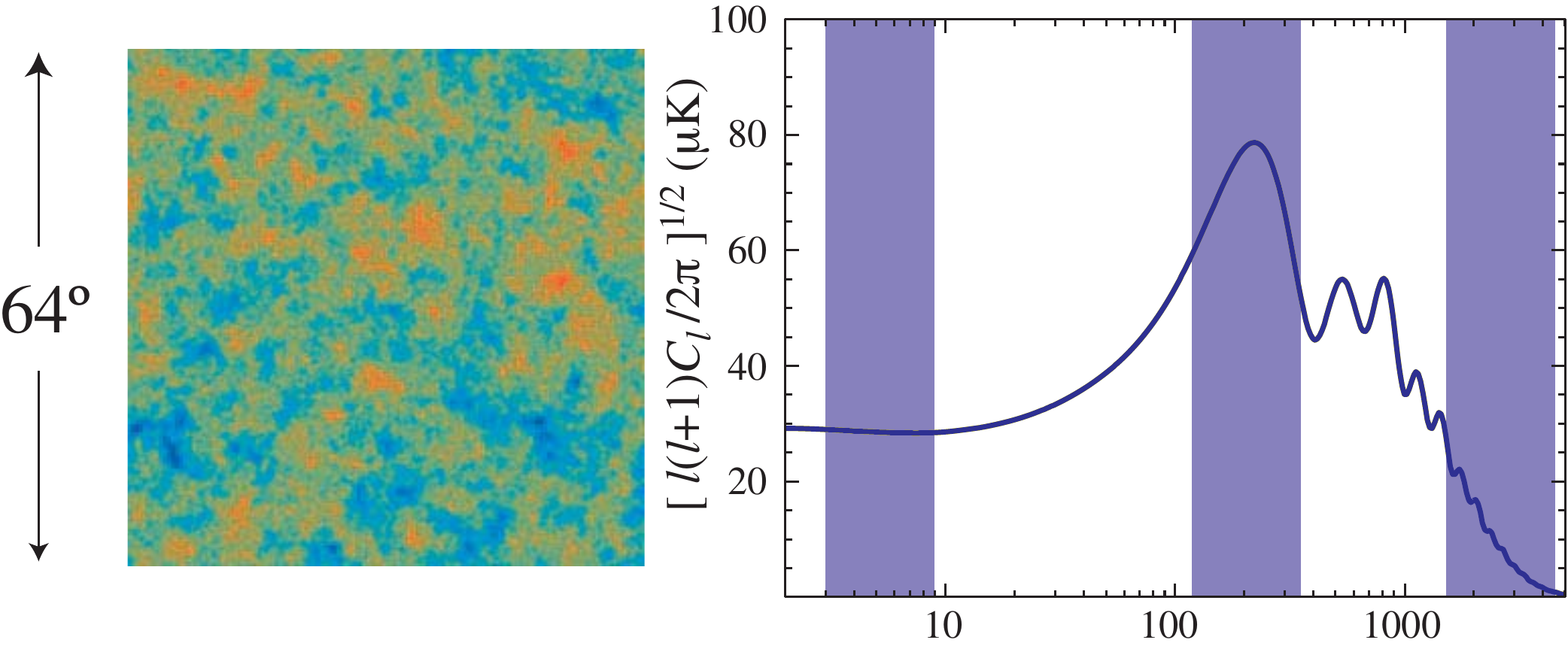}
\includegraphics[width=4.5in]{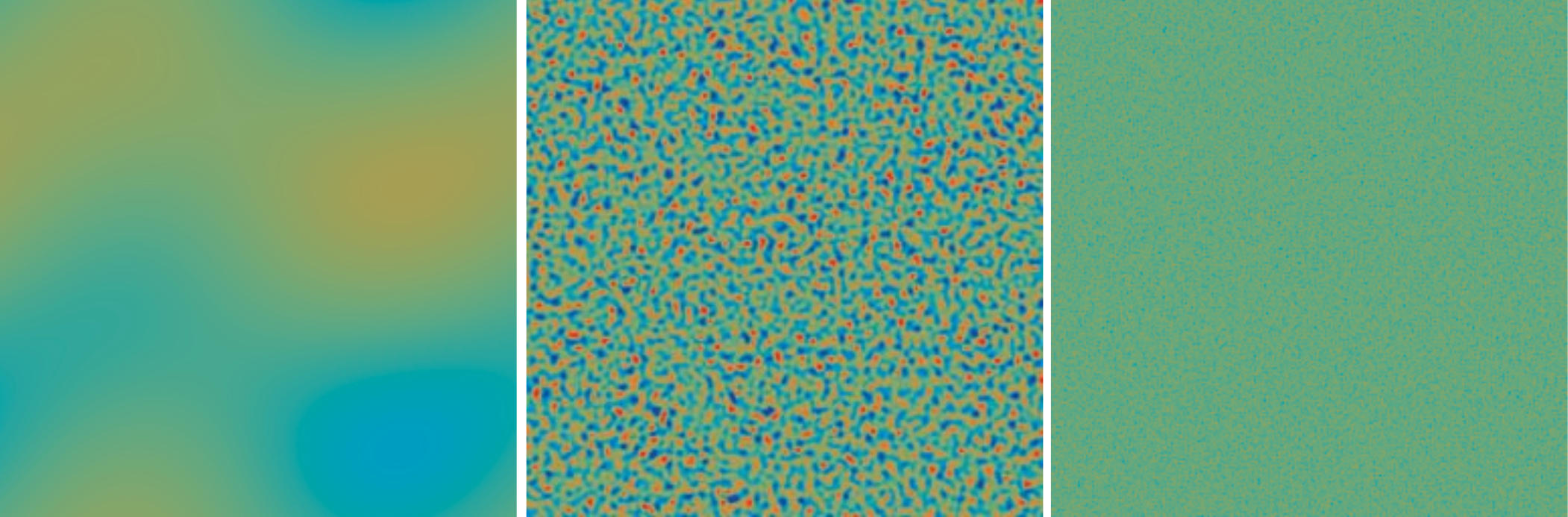}
\caption{From temperature maps to power spectrum.  The original temperature fluctuation map
(top left) corresponding to a simulation of the power spectrum (top right) can be band filtered
to illustrate the power spectrum in three characteristic regimes: the large-scale gravitational 
regime of COBE, the first acoustic peak where most of the power lies, and the damping tail where
fluctuations are dissipated.  Adapted from \cite{HuWhiSciAm}.} \label{fig:filter}
\end{center}
\end{figure}

\section{Temperature Anisotropy from Recombination}
\label{sec:anisotropy}

Spatial variations in the CMB temperature at recombination are seen as temperature
anisotropy by the observer today.
The temperature anisotropy of the CMB was first detected in 1992 by the COBE
DMR instrument (\citealt{Smoetal92}).   These corresponded to variations of order
$\Delta T/T \sim 10^{-5}$ across $10^{\circ}-90^{\circ}$ on the sky (see Fig.~\ref{fig:filter}). 

\begin{figure}[tb]
\begin{center}
\includegraphics[width=4in]{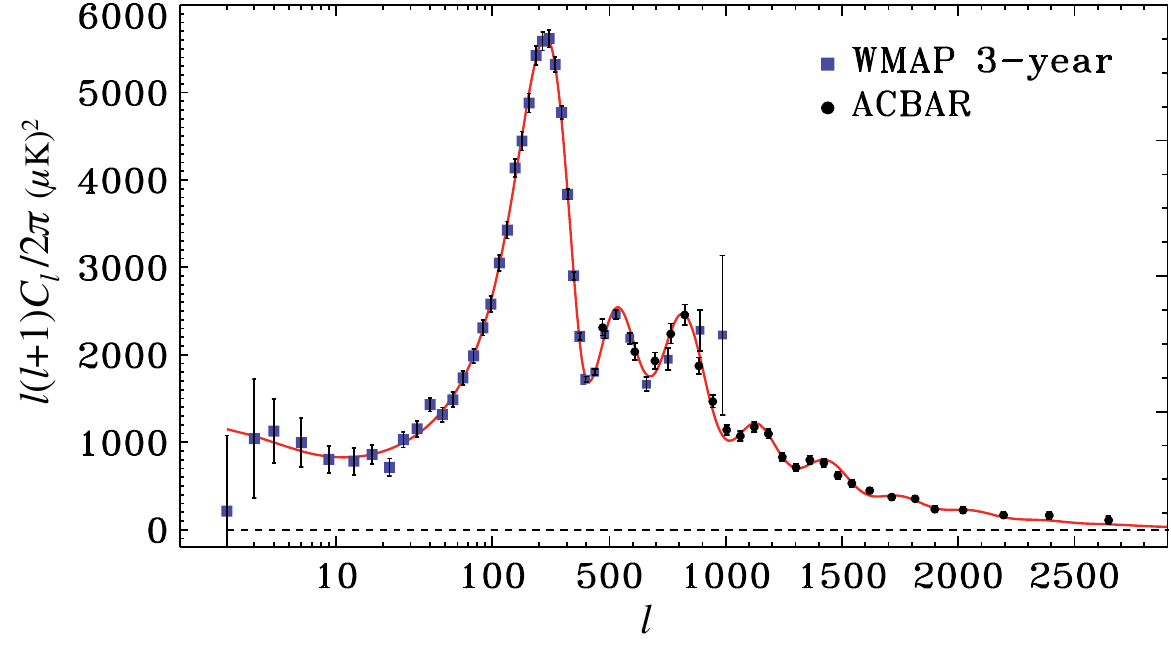}
\caption{Temperature power spectrum from recent measurements from WMAP and 
ACBAR along with the best fit $\Lambda$CDM model.  The main features of the temperature
power spectrum including the first 5 acoustic peaks and damping tail have now been measured.
 Adapted from \cite{Reietal08}.}. \label{fig:cldata}
\end{center}
\end{figure}

 Most of the structure in the temperature anisotropy however is
associated with acoustic oscillations of the photon-baryon plasma on $\sim 1^{\circ}$ scales. 
Throughout the 1990's constraints on the location of the first peak
steadily improved culminating with the determinations of
the TOCO (\citealt{Miletal99}), Boomerang, (\citealt{deBetal00}) and Maxima-1 
(\citealt{Hanetal00}) experiments.  Currently from WMAP (\citealt{Speetal06})
and ground based experiments,
we have precise measurements of the first five acoustic peaks (see Fig.~\ref{fig:cldata}). 
Primary fluctuations beyond this scale (below $\sim 10'$) are damped by Silk damping (\citealt{Sil68}) as verified
observationally first by the CBI experiment (\citealt{Padetal01}).

In this section, we deconstruct the basic physics behind these phenomena.  We begin with
the geometric projection of temperature inhomogeneities at recombination onto the sky of
the observer in \S\ref{sec:projection}.    We continue with the basic equations of fluid
mechanics that govern acoustic phenomena in \S\ref{sec:basics}.   We add gravitational
(\S\ref{sec:gravity}), baryonic (\S\ref{sec:baryons}),
 matter-radiation (\S\ref{sec:matter}),  and dissipational (\S\ref{sec:damping}) effects
 in the sections that follow.   Finally we put these pieces back  together to discuss the information
content of the acoustic peaks in  \S \ref{sec:information}.

\subsection{Anisotropy from Inhomogeneity}
\label{sec:projection}

Given that the CMB radiation is blackbody to experimental accuracy (see Fig.~\ref{fig:firas}),
one can
characterize its spatial and angular distribution by its temperature at   the position
$\bx$ of the observer in the direction $\bn$ on the observer's sky
\myequation{{f(\nu,\bn,\bx)} = [\exp(2\pi \nu/{T(\bn;\bx)}-1]^{-1}\,,}
where 
 $\nu=E/2\pi$ is the observation frequency.
The hypothetical observer could be an electron in the intergalactic medium  or the
true observer on earth or L2.  When the latter is implicitly meant, we will take
$\bx ={\bf 0}$.  We will occasionally suppress the coordinate $\bx$ when this position is to be understood.

For statistically isotropic, Gaussian random temperature fluctuations a harmonic
description is more efficient than a  real space description.  For the
 angular structure at the position of the observer, the appropriate harmonics
 are the
spherical harmonics.   These are the eigenfunctions of the Laplace operator on the sphere
and form a complete basis for scalar functions on the sky
\myequation{\Theta(\bn) = {T(\bn) -\bar T \over \bar T}
 = \sum_{\ell m} \Theta_{\ell m} Y_{\ell m}(\bn)\,.}

\begin{figure}[tb]
\begin{center}
\includegraphics[width=2.5in]{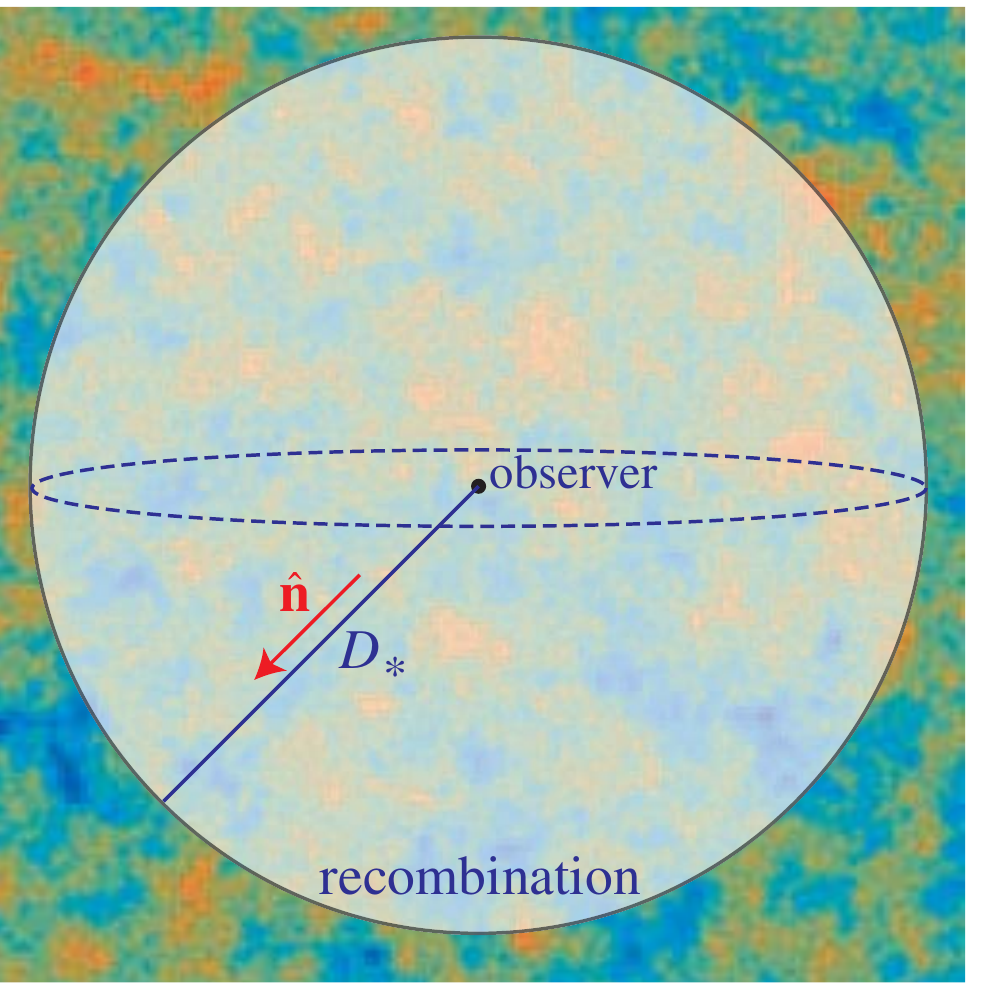}
\caption{From inhomogeneity to anisotropy.  Temperature inhomogeneities at recombination
are viewed at a distance $D_*$ as anisotropy on the observer's sky.} \label{fig:projection}
\end{center}
\end{figure}

{For  {statistically isotropic}
 fluctuations, the ensemble average of the temperature
		fluctuations are described by the {power spectrum}}
\myequation{ \langle {\Theta_{\ell m}}^* {\Theta_{\ell' m'}} \rangle = \delta_{\ell \ell'} \delta_{m m'} C_\ell \,. \label{eqn:cl}}
Moreover, the power spectrum contains all of the statistical information in the field
if the fluctuations are Gaussian.
{Here $C_{\ell}$ is dimensionless but is often shown with
 units of squared temperature, \emph{e.g.} {$\mu$K$^2$}, by multiplying through by the
background temperature today  $\bar T$.   The correspondence between angular
size and amplitude of fluctuations and the power spectrum is shown in Fig.~\ref{fig:filter}.

Let us begin with the simple approximation 
that the temperature field at recombination is isotropic but inhomogeneous
and the anisotropy viewed
 at the present is due to the observer seeing different portions of the recombination
 surface in different directions (see Fig.~\ref{fig:projection}).  We will develop this picture
further in the following sections with gravitational redshifts,
dipole or Doppler anisotropic sources, the
finite duration of recombination, and polarization.

Under this simple instantaneous recombination approximation, the angular temperature fluctuation
distribution is simply a projection of the spatial temperature fluctuation
\myequation{{\Theta(\bn)} = \int d D\, {\Theta(\bx)} \delta(D - D_*)\,,
\label{eqn:thetadelta}}
{where $D=\int dz/H$ is the {comoving distance} and $D_*$ denotes the 
distance a CMB photon travels from
recombination.}  Here $\Theta(\bx)=(T(\bx)-\bar T)/\bar T$ is the spatial temperature fluctuation
at recombination.  Note that the cosmological redshift does not appear in the temperature
fluctuation since the background and fluctuation redshift alike.

The spatial power spectrum at recombination can be likewise be described
by its harmonic modes.  In a flat geometry, these are Fourier modes
\myequation{ \Theta{(\bx)} = \int {d^3 k \over (2\pi)^3} {\Theta(\bk)} e^{i \bk \cdot \bx} \,.}
More generally they are the eigenfunctions of the Laplace operator on the three dimensional
space with constant curvature of a general Fried\-mann-Robert\-son-Wal\-ker metric
(see \emph{e.g.} \citealt{Hu04a}).
For a statistically homogeneous spatial distribution, the two point function is described
by the power spectrum
\myequation{ \langle \Theta(\bk)^* \Theta(\bk') \rangle = (2\pi)^3 \delta(\bk - \bk') {P(k)}\,. 
\label{eqn:pk}}

These relations tell us that the amplitude of the angular and spatial power spectra 
are related.  A useful way of establishing this relation and quantifying the power 
in general is to describe the contribution per
logarithmic interval to the variance of the respective fields:
\begin{eqnarray}
 \langle {\Theta(\bx) \Theta(\bx)} \rangle = \int {d^3 k \over (2\pi)^3 }P(k)  =
  \int {d\ln k}\, {k^3 P(k) \over 2\pi^2} \equiv \int {d \ln k}\, {\Delta_{T}^2(k)} \,.
  \label{eqn:logpower}
\end{eqnarray}
A scale invariant spectrum has equal contribution to the variance per e-fold
$\Delta_{T}^{2}=k^{3}P(k)/2\pi^{2}=$ const.
To relate this to the amplitude of the angular power spectrum, we expand equation (\ref{eqn:thetadelta})
in Fourier modes
\myequation{{\Theta(\bn)} = \int {d^3 k \over (2\pi)^3} {\Theta(\bk)} e^{i \bk \cdot D_* \bn}\,. }
The Fourier modes themselves can be expanded in spherical harmonics with the relation
\myequation{ {e^{i \bk D_* \cdot \bn}} = 4\pi \sum_{\ell m} i^{\ell} {j_\ell}(k D_*) Y_{\ell m}^*({\hat \bk}) Y_{\ell m}(\bn)\,,}
where $j_{\ell}$ is the spherical Bessel function.
Extracting the multipole moments, we obtain
\myequation{ {\Theta_{\ell m}} = \int {d^3 k \over (2\pi)^3} {\Theta(\bk)} 4\pi i^\ell j_{\ell}(k D_*) Y_{\ell m}(\bk)\,. }
We can then relate the angular and spatial two point functions (\ref{eqn:cl}) and (\ref{eqn:pk})
\begin{eqnarray}
 \langle {\Theta_{\ell m}^* \Theta_{\ell' m'}}\rangle 
		= \delta_{\ell\ell'} \delta_{m m'} 4\pi \int d\ln k\, j_\ell^2(k D_*){\Delta_{T}^2(k)}
		= \delta_{\ell\ell'} \delta_{m m'} C_{\ell}\,.
		\label{eqn:thetalm}
		\end{eqnarray}
Given a slowly varying, nearly scale invariant spatial power spectrum we can take $\Delta_{T}^{2}$
out of the integral and evaluate it at the peak of the Bessel function $k D_{*} \approx \ell$. 
The remaining integral can be evaluated in closed form
 $\int_0^\infty {j_\ell^2(x)} d\ln x= 1/[2{\ell(\ell+1)}]$ yielding the final result
\myequation{{C_\ell} \approx  {2\pi \over \ell(\ell+1) }
\Delta_{T}^2(\ell/D_*)\,.
\label{eqn:cldelta}}
Likewise even slowly varying features like the acoustic peaks of \S\ref{sec:basics} also
mainly map to multipoles of $\ell \approx k D_*$ due to the delta function like behavior
of $j_\ell^2$.  

It is therefore common  to plot the angular power spectrum as
\begin{equation}
{\cal C}_{\ell} \equiv {\ell(\ell+1) \over 2\pi} C_\ell  \approx \Delta_{T}^2 \,.
\end{equation}
It is also common to plot ${\cal C}_{\ell}^{1/2} \approx \Delta_{T}$, the logarithmic 
contribution to the rms of the
field, in units of $\mu$K.

Now let us compare this expression to the  {variance per log interval} in multipole space:}
\begin{eqnarray}
 \langle {T(\bn) T(\bn)} \rangle &=&\sum_{\ell m} \sum_{\ell' m'}
\langle { T_{\ell m}^* T_{\ell' m'}} \rangle Y_{\ell m}^*(\bn) Y_{\ell' m'}(\bn)\nonumber\\
&=& \sum_{\ell}{ C_{\ell}}\sum_{m} Y_{\ell m}^*(\bn) Y_{\ell m}(\bn) =  \sum_{\ell} { {2 \ell +1 \over 4\pi} C_{\ell}}\,.
\end{eqnarray}
For variance contributions from $\ell \gg 1$,
\begin{equation}
 \sum_{\ell} { {2 \ell +1 \over 4\pi} C_{\ell}} \approx
 \int d\ln \ell  { {\ell(2 \ell +1) \over 4\pi}} C_{\ell} 
\approx 
  \int d\ln \ell  { {\ell(\ell +1) \over 2\pi}} C_{\ell} \,.
\end{equation}
Thus ${\cal C}_{\ell}$ is also approximately the variance per log interval in angular
space as well.

\subsection{Acoustic Oscillation Basics}
\label{sec:basics}

\emph{Thomson Tight Coupling ---}
To understand the angular pattern of temperature fluctuations seen by the observer
today, we must understand the spatial temperature pattern at recombination. 
That in turn requires an understanding of the dominant physical processes in
the plasma before recombination.

{{Thomson scattering} of photons off of free electrons is the most important process given
its relatively large cross section (averaged over polarization states) }
\begin{eqnarray}
{ \sigma_T }   & = & {8\pi \alpha^2 \over 3 m_e^2} = 6.65 \times 10^{-25} {\rm cm}^{2} \,.
\end{eqnarray}

The important quantity to consider is the mean free path of a photon given Thomson
scattering and a medium with a free electron density $n_{e}$.  Before recombination
when the ionization fraction $x_{e} \approx 1$ this density is given by 
\begin{eqnarray}
{n_e} &=& (1-Y_p) x_e n_b \nonumber  \\
&=&1.12 \times 10^{-5} (1-Y_p) x_e  \Omega_b h^2 {(1+z)^3} {\rm cm}^{-3}\,.
\end{eqnarray}
  The comoving mean free path
$\lambda_{C}$ is given by 
\begin{eqnarray}
\lambda_{C}^{{-1}} \equiv \dot \tau  \equiv n_e \sigma_T a \,,
\end{eqnarray}
{where  the extra factor of $a$ comes from converting physical to comoving coordinates and
$Y_p$ is the primordial helium mass fraction}.
We have also represented this mean free path in terms of an scattering absorption coefficient
 $\dot\tau$ where
dots are conformal time $\eta \equiv \int dt /a$ derivatives and 
$\tau$ is the optical depth.

{Near {recombination} ($z \approx 10^3$, $x_e \approx 1$) and given 
$\Omega_b h^2 \approx 0.02$, and $Y_p = 0.24$, the mean free path is}
\begin{eqnarray}
{\lambda_C} \equiv \frac{1}{\dot \tau} \sim 2.5 {\rm Mpc}\,.
\label{eqn:mfp}
\end{eqnarray}
This scale is almost two orders of magnitude smaller than the horizon at recombination. 
{On scales $\lambda \gg \lambda_C$ photons are {tightly coupled} to the electrons
by Thomson scattering which in turn are tightly coupled to the baryons by Coulomb interactions.}

As a consequence, any bulk motion of the photons must be shared by the baryons.
In fluid language, the two species have a single bulk velocity 
		$v_\gamma = v_b$ and hence no entropy generation or heat conduction occurs.  
		Furthermore, the shear viscosity of the fluid is negligible.  
		Shear viscosity is related to anisotropy in the radiative pressure or stress
		and rapid scattering isotropizes the photon distribution.  
		This is also the reason why in equation~(\ref{eqn:thetadelta}) we took the
		photon distribution to be isotropic but inhomogeneous.

		We shall see that
		the fluid motion corrects this by allowing dipole $\ell=1$
		 anisotropy in the distribution but no  higher $\ell$ modes.
		It is only on scales smaller than the diffusion scale that
		 radiative viscosity ($\ell=2$)
		  and heat conduction
		 becomes sufficiently large to dissipates the bulk motions of the plasma
		 (see \S\ref{sec:damping}).

\medskip\noindent\emph{Zeroth Order Approximation ---}
To understand the basic physical picture,
let us begin our discussion of acoustic oscillations in the tight coupling regime with
a simplified system which we will refine as we progress.

First let us ignore the dynamical impact of the baryons on the fluid motion.
Given that the baryon and photon velocity are equal and the momentum 
density of a relativistic fluid is given by
$(\rho+p)v$, where $p$ is the pressure, and $v$ is the fluid velocity, 
 this approximation relates to the quantity
\begin{eqnarray}
 {R} & \equiv & {(\rho_b + p_b )v_b \over ( \rho_\gamma + p_\gamma ) v_\gamma }  = {\rho_b + p_b \over \rho_\gamma + p_\gamma} ={ {3 \rho_b \over 4\rho_\gamma}} \nonumber\\
     & \approx & {0.6} \left( {\Omega_b h^2 \over 0.02} \right) \left( {a \over 10^{-3}} \right) \,,
     \label{eqn:Rfactor}
\end{eqnarray}
where we have used the fact that 
$\rho_\gamma \propto T^4$ so that its value is fixed by the redshifting background
 $\bar T = 2.725(1+z)$K.   Neglect of the baryon inertia and momentum
  only fails right around recombination.
 
Next, we shall assume that the background expansion is matter dominated to relate
time and scale factor.  The validity of this approximation depends on the matter-radiation
ratio
\myequation{ {\rho_m \over \rho_r} = {3.6} \left( {\Omega_m h^2 \over 0.15} \right)
	\left( {a \over 10^{-3}} \right) \,, \label{eqn:matterradiation}}
and is approximately valid during recombination and afterwords.   One expects from these
arguments that order unit differences between the real universe and our basic description 
will occur.  We will in fact use these differences in the following sections to show how
the baryon and matter densities are measured from the acoustic peak morphology.

Finally, we shall consider the effect of pressure forces and neglect gravitational forces.
While this is not a valid approximation in and of itself,
 we shall see that for a photon-dominated system, the error in ignoring gravitational forces exactly cancels with that from ignoring gravitational redshifts that photons experience
after recombination (see \S\ref{sec:gravity}).

\medskip\noindent\emph{Continuity Equation ---}
Given that Thomson scattering neither creates nor destroys photons, the continuity
equation implies that the photon number density only changes due to flows into and
out of the volume.  In a non expanding universe that would require
\myequation{
{\dot n_\gamma} + \nabla \cdot ({n_\gamma {\bf v}_\gamma}) = 0\,. }
Since $n_\gamma$ is the number density of photons per unit physical (not comoving)
volume, this equation must be corrected for the expansion.   The effect of the
expansion can alternately be viewed as that of the Hubble flow diluting the
number density everywhere in space.   Because number densities scale as
 $n_\gamma \propto a^{-3}$, the expansion alters the continuity equation as
\myequation{
{\dot n_\gamma} + 3 {n_\gamma} {\dot a \over a} + \nabla \cdot ({n_\gamma {\bf v}_\gamma})= 0\,.\label{eqn:cont2}}
Since we are interested in small fluctuations around the background, let us linearize
the equations $n_{\gamma} \approx \bar n_{\gamma} + \delta n_{\gamma}$
and drop terms that are higher than first 
order in $\delta n_{\gamma}/n_{\gamma}$ and $v_{\gamma}$.  
Note that $v_{\gamma}$ is first order in the number density fluctuations since
as we shall see in the Euler equation discussion below it is generated from the  pressure gradients associated with density fluctuations.

The continuity equation (\ref{eqn:cont2})
for the fluctuations becomes
\begin{eqnarray}
\left( { {\delta n_\gamma \over n_\gamma}}
  \right)^{\cdot} &=& - \nabla \cdot { {\bf v}_\gamma} \,.
\end{eqnarray}
Since the 
{{number density}
 $n_\gamma \propto T^3$, the fractional density fluctuation is
 related to the {temperature fluctuation} $\Theta$ as}
\myequation{ {\delta n_\gamma \over n_\gamma} = 3 {{\delta T \over T}} \equiv 3 {\Theta} \,.}
Expressing the continuity equation in terms of $\Theta$, we obtain
\myequation{
{ \dot \Theta}  = -{1 \over 3} \nabla \cdot { {\bf v}_\gamma} \,.}
Fourier transforming this equation, we get
\myequation{
{ \dot \Theta}  = -{1 \over 3} {i \bf k} \cdot { {\bf v}_\gamma}\,,
\label{eqn:continuity0}}
for the relationship between the Fourier mode 
amplitudes.

\medskip\noindent\emph{Euler Equation ---}
Now let us examine the origin of the fluid velocity.   In the background, the
velocity vanishes due to isotropy.  However, Newtonian mechanics dictates
that pressure forces will generate particle momentum as 
{ ${ \dot {\bf q}}  = {\bf F}$.}
Newton's law must also be modified for the expansion.   If we associate
 the de Broglie wavelength with the inverse momentum,
 this wavelength
also stretches with the expansion.   For photons, this accounts for the
redshift factor.  For non-relativistic matter, this means that bulk velocities
decay with the expansion.   In either case, we generalize Newton's law
to read
\myequation{ { \dot {\bf q}} + {\dot a \over a} {\bf q} ={\bf F}\,.}

For a collection of particles, the relevant quantity is the momentum density
\begin{equation}
(\rho_\gamma + p_\gamma) {{\bf v}_\gamma}
\equiv \int {d^3 q \over (2\pi)^3} {\bf q} f\,,
\end{equation}
and likewise the force becomes
a force density.  For the photon-baryon fluid, this force density is provided by
the pressure gradient.  The result is the Euler equation
\begin{eqnarray}
[(\rho_\gamma + p_\gamma){{\bf v}_\gamma} ]^{\cdot} =
- 4 {\dot a \over a}(\rho_\gamma + p_\gamma){ {\bf v}_\gamma } - { \nabla p_\gamma} \,,
\end{eqnarray}
where the 4 on the right hand side  comes from combining the redshifting of wavelengths
with that of number  densities.
Since photons have an equation of state $p_{\gamma} = \rho_{\gamma}/3$, the Euler
equation becomes
\begin{eqnarray}
{4 \over 3}\rho_\gamma { \dot {\bf v}_\gamma} &=&- {1 \over 3 } {\nabla \rho_\gamma}\,,
\nonumber\\
{ \dot {\bf v}_\gamma} &=& - { \nabla \Theta} \,.
\end{eqnarray}

In Fourier space, the Euler equation becomes
\myequation{
{\dot {\bf v}_\gamma}  = -i \bk {\Theta} \,.}
The factor of $i$ here represents the fact that the temperature
maxima and minima are zeros of the velocity in real space due to the gradient
relation, \emph{i.e.} temperature and velocity have a  $\pi/2$ phase shift.
It is convenient therefore to define the velocity amplitude to absorb this
factor
\myequation{ {{\bf v}_\gamma} \equiv -i {v_\gamma} \hat{\bf k}\,.}
The direction of the fluid velocity is always parallel to the wavevector $\bk$ for
linear (scalar) perturbations and
hence we can write the Euler equation as
\myequation{{\dot v_\gamma} = k{\Theta}\,.
\label{eqn:euler0}}

\begin{figure}[tb]
\begin{center}
\includegraphics[width=5in]{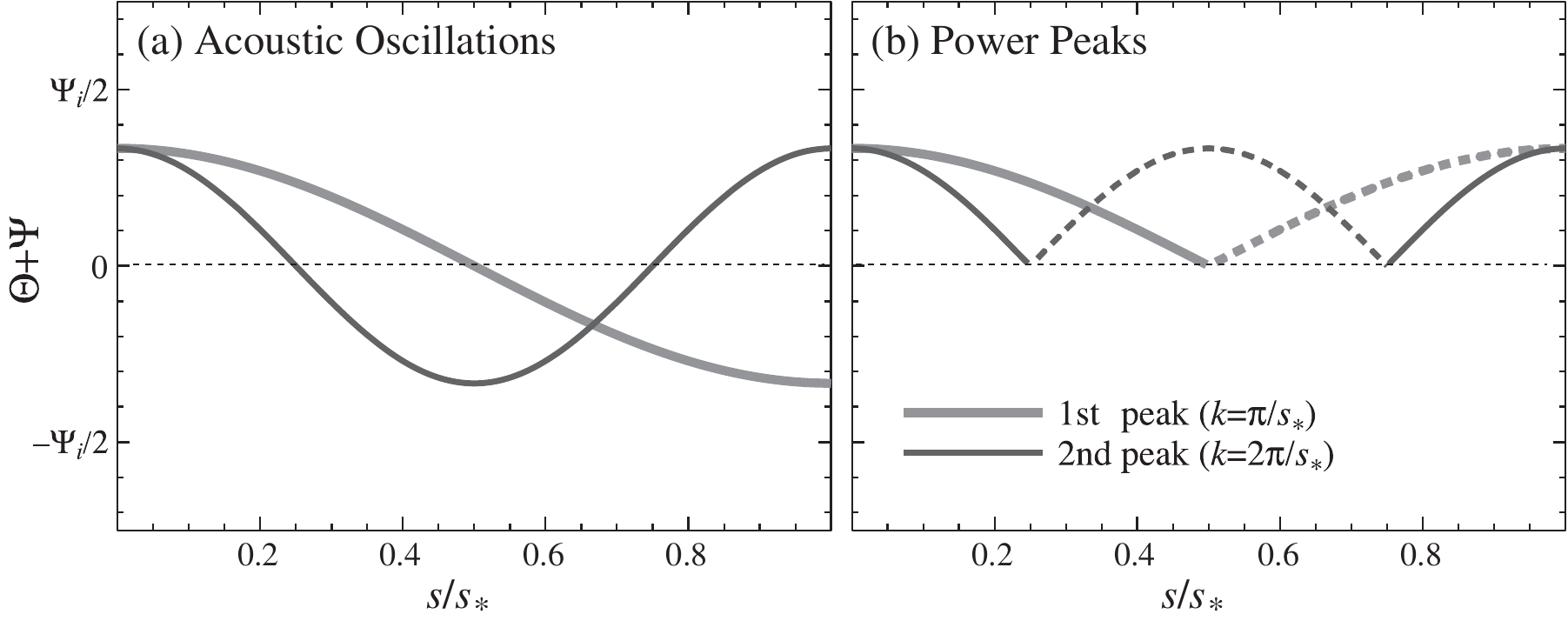}
\caption{Acoustic oscillation basics. All modes start from the same initial epoch with time denoted
by the sound horizon relative to the sound horizon at recombination $s_*$.   (a) Wavenumbers
that reach extrema in their effective temperature $\Theta+\Psi$ (accounting for gravitational
redshifts \S\ref{sec:gravity}) at $s_*$ form a harmonic series $k_n= n \pi/s_*$.  (b)
Amplitude of the fluctuations is the same for the maxima and minima without baryon inertia.
Adapted from \cite{HuDod02}.} \label{fig:osc1}
\end{center}
\end{figure}

\medskip\noindent\emph{Acoustic Peaks ---}
Combining the continuity (\ref{eqn:continuity0}) and Euler
(\ref{eqn:euler0}) equations to eliminate the fluid velocity, we get the {simple harmonic oscillator} equation
\myequation{{\ddot \Theta} + {c_s^2} k^2 {\Theta} = 0\,,}
{where the adiabatic sound speed $c_{s}^{2}=1/3$ for the photon-dominated fluid and
more generally is defined as}
\myequation{ {c_s^2} \equiv \frac{{\dot p_\gamma}}{ {\dot \rho_\gamma}}\,.}

The solution to the oscillator equation can be specified given two initial
conditions $\Theta(0)$ and $v_{\gamma}(0)$ or $\dot\Theta(0)$,
\myequation{{\Theta(\eta)} = {\Theta(0)} \cos(k {s})+ {{\dot \Theta(0)} \over k c_s} 
\sin(k {s})\,,\label{eqn:oscillatorsolution}}
{where the {sound horizon} is defined as 
\begin{equation}
{s} \equiv \int {c_s} d\eta \,.
\end{equation}
In real space, these oscillations appear as standing waves for each Fourier mode.

These standing waves continue to oscillate until recombination.  At this point the
free electron density drops drastically (see Fig.~\ref{fig:recombination})
 and the photons freely stream to the observer.
The pattern of acoustic oscillations on the recombination surface seen
by the observer becomes the acoustic peaks in the temperature anisotropy.

Let us focus on the adiabatic mode which starts with a finite density or temperature
fluctuation and vanishing velocity perturbation.  At recombination $\eta_{*}$, the
oscillation reaches (see Fig.~\ref{fig:osc1})
\myequation{{\Theta(\eta_*)} = {\Theta(0)} \cos({k} {s_*})\,.}
Considering a spectrum of $k$ modes, the critical feature of these oscillations
are that they are temporally coherent.   The underlying assumption is that
 fluctuations of all wavelengths originated at $\eta=0$ or at least $\eta \ll \eta_*$.  
 Without inflation this would
 violate causality for long wavelength fluctuations, \emph{i.e.} the analogue of the horizon problem
 for perturbations.    With inflation, superhorizon 
 modes originate during an inflationary epoch $\eta_i \ll \eta_*$.

Modes caught in the {extrema} of their oscillation follow a harmonic
relation
\myequation{ {k_n} s_* = {n}\pi\,, \quad n=1,2,3 \ldots}
{yielding a {fundamental scale} or frequency, related to the inverse
{sound horizon}}
\myequation{ {k_A} = \pi/s_*\,.} 
Since
the power spectrum is proportional to the square of the fluctuation, both
maxima and minima contribute peaks in the spectrum.  Observational verification
of this harmonic series is the primary evidence for inflationary adiabatic initial
conditions (\citealt{HuWhi96c}).

{The fundamental {physical scale} is translated into a fundamental {angular
		scale} by simple projection according to the angular diameter distance $D_A$}
\eqnsize\begin{eqnarray}
{\theta_A}  &=& {\lambda_A}/D_A\,, \nonumber \\
{\ell_A}       &=& {k_A} D_A \,,
\label{eqn:thetaA}
\end{eqnarray}
(see Eqn.~(\ref{eqn:cldelta})).
{In a flat universe, the distance is simply $D_A=D \equiv \eta_0 - \eta_* \approx {\eta_0}$,
the horizon distance, and ${k_A} = \pi /s_* = \sqrt{3}\pi / \eta_*$  so }
\myequation{{\theta_A} \approx {\frac{\eta_*}{\eta_0}}\,.}
{Furthermore, in a {matter-dominated} universe $\eta \propto a^{1/2}$ so $\theta_A
\approx 1/30 \approx 2^\circ$ or}
\myequation{{\ell_A} \approx {200}\,.}
We shall see in \S\ref{sec:information} that radiation and dark energy 
introduce important corrections to this prediction from their influence on $\eta_{*}$
and $D_{*}$ respectively.   Nonetheless it is remarkable that this simple argument predicts the
basics of acoustic oscillations: their existence, coherence and fundamental scale.

\begin{figure}[tb]
\begin{center}
\includegraphics[width=3.5in]{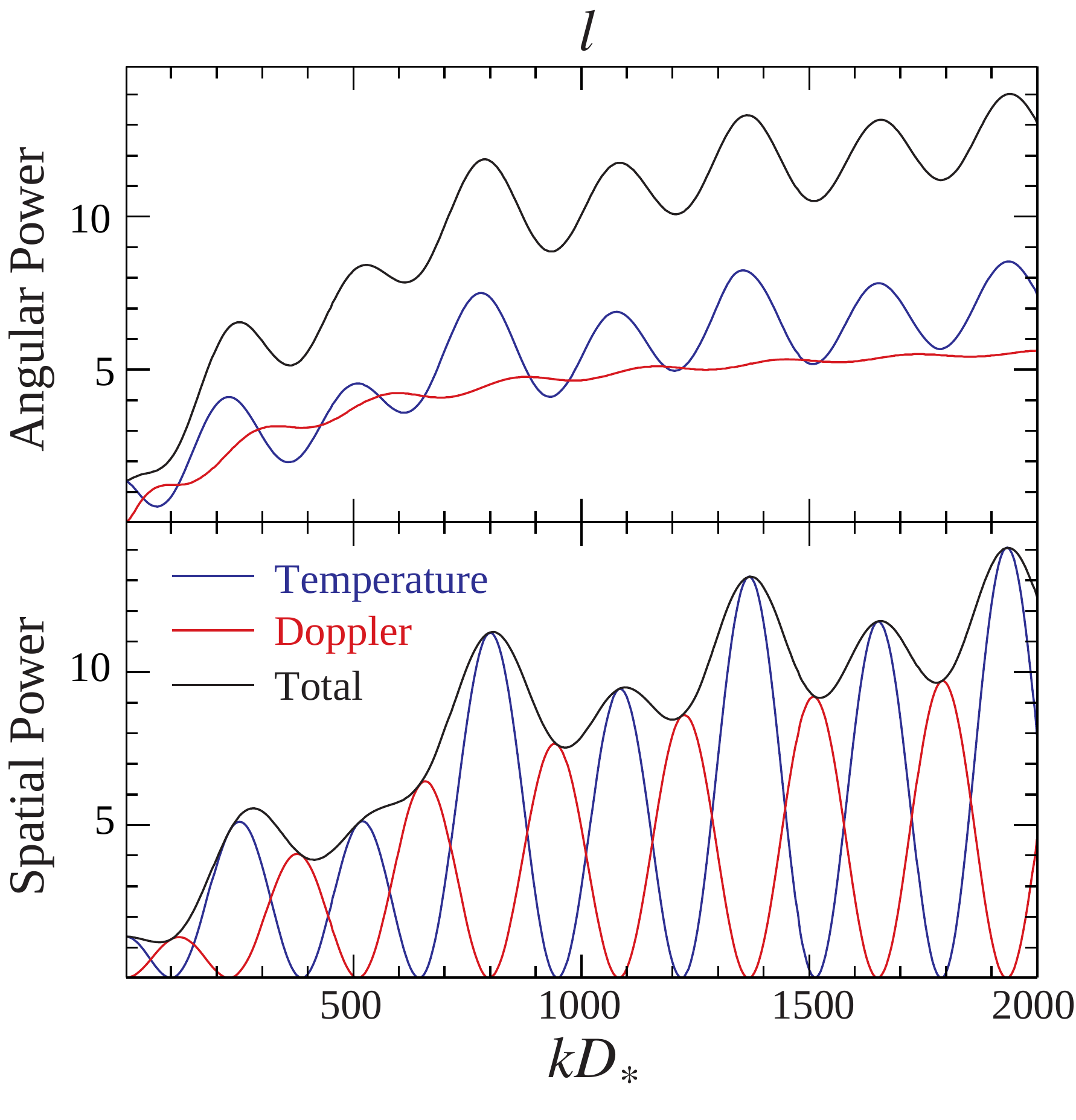}
\caption{Doppler effect. The Doppler effect provides fluctuations of
comparable strength to the local temperature fluctuations from acoustic oscillations in $k$-space (lower) providing features at the troughs of the latter.  In angular space, projection
effects smooth the Doppler features leaving an acoustic morphology that
reflects the temperature oscillations.  The peak height modulation comes from
the baryon inertia (\S \ref{sec:baryons}) and the gradual increase in power with $\ell$
from radiation domination (\S \ref{sec:matter}).} \label{fig:dop}
\end{center}
\end{figure}

\medskip\noindent\emph{Acoustic Troughs: Doppler Effect ---}
Acoustic oscillations also imply that the plasma is moving relative to the observer.
This bulk motion imprints a temperature anisotropy via the Doppler effect
\myequation{ \left( {\Delta T \over T} \right)_{\rm dop} = \bn \cdot {{\bf v}_\gamma} \,.}
{Averaged over directions}
\myequation{ \left( {\Delta T \over T} \right)_{\rm rms} = {{v_\gamma} \over \sqrt{3} } \,,}
which given the acoustic solution of Eqn.~(\ref{eqn:oscillatorsolution}) implies
\begin{eqnarray}
 {{v_\gamma} \over \sqrt{3} } & =& - {\sqrt{3} \over k} { \dot \Theta} = {\sqrt{3} \over k} k c_s \, \Theta(0)  {\sin(k s)} \nonumber\\
&=& \Theta(0) {\sin(k s)} \,.
\end{eqnarray}

Interestingly, the 
{{Doppler effect} for the photon-dominated system is of {equal amplitude}
and $\pi/2$ {out of phase}: extrema of temperature are turning points of velocity}.
If we simply add the $k$-space temperature and Doppler effects in quadrature we would obtain
\myequation{ \left( {\Delta T \over T} \right)^2 = \Theta^2(0) [\cos^2 (k s) + \sin^2(k s)] = {\Theta^2(0)} \,.}
In other words there are no preferred $k$-modes at harmonics and a scale invariant initial temperature
spectrum would lead to a scale invariant spatial power spectrum at recombination.  However the Doppler effect carries an angular dependence
that changes its {projection} on the sky $\bn \cdot {\bf v_\gamma} \propto \bn \cdot \hat{\bf k}$}.
In a coordinate system where 
 $\hat{\bf z} \parallel \hat{\bf k}$, this angular dependence yields an extra factor of
 $Y_{10}$  in the analogue of Eqn.~(\ref{eqn:thetalm}).  This extra factor can 
 be reabsorbed into the total angular dependence through Clebsch-Gordan recoupling
 (\citealt{HuWhi97a})
\myequation{ {Y_{10}} Y_{\ell 0} \rightarrow {Y_{\ell \pm 1 \, 0}}\,.}
The  recoupling implied for the radial harmonics changes $j_\ell(x) \rightarrow j_\ell'(x)$.
The projection kernel $j_\ell'(x)$ lacks a strong feature at $\ell \sim x$ and so Doppler contributions
in $k$ are spread out in $\ell$.   This is simply a mathematical way of stating that the
Doppler effect vanishes when the observer is looking perpendicular to ${\bf v} \parallel {\bf k}$ whereas
it is in that direction that the acoustic peaks in temperature gain most of their contribution.
The net effect, including baryonic effects that we discuss below, is that the peak
structure is dominated by the local temperature at recombination and not the local fluid motion.

\subsection{Gravito-Acoustic Oscillations}
\label{sec:gravity}

Thus far we have neglected gravitational forces and redshifts in our discussion of plasma motion.  The true system exhibits gravito-acoustic or Jeans oscillations.  
We were able to employ this swindle to get the basic properties of acoustic oscillations
because in a photon-dominated plasma the effect of gravitational forces and gravitational
redshifts exactly cancel in  constant gravitational potentials.   To go beyond the
photon dominated plasma and matter dominated expansion approximation, we now
need to include these effects.   Furthermore, we shall see that the gravitational potential
perturbations from inflation are also the source of the initial temperature fluctuation.  

\medskip\noindent\emph{Continuity Equation and Newtonian Curvature ---}
The photon continuity equation for the number density is altered by gravity since the presence of a gravitational
potential alters the coordinate volume.  Formally in general relativity this comes from
the space-space piece of the metric --  a spatial {curvature perturbation} $\Phi$:
\begin{equation}
ds^{2} = a^{2}[-(1+2\Psi)d\eta^{2} + (1+2\Phi)dx^{2}]
\label{eqn:lineelement}
\end{equation}
for a flat cosmology.

We can think of this curvature perturbation as changing the local scale factor
$a \rightarrow a(1+\Phi)$ so that
the expansion dilution is generalized to
\myequation{ {\dot a \over a} \rightarrow {\dot a \over a} + {\dot \Phi}\,.}
Hence the full continuity equation is now given by
\begin{eqnarray}
({\delta n_\gamma})^{\cdot} &=& - 3{ \delta n_\gamma} 
 {\dot a \over a} - 3 n_\gamma { \dot \Phi }
- n_\gamma \nabla \cdot {{\bf v}_\gamma}\,,
\end{eqnarray}
{or}
\myequation{{\dot \Theta} = -\frac{1}{3} k {v_\gamma} -{\dot\Phi}\,.
\label{eqn:continuitywithgravity}}

\medskip\noindent\emph{Euler Equation and Newtonian Forces ---}
Likewise the gravitational force from gradients in the gravitational 
potential ($\Psi$, formally the time-time piece of the metric perturbation) modifies
the momentum conservation equation.   The Newtonian force ${\bf F} = -m \nabla \Psi$ generalized
to momentum density brings the {Euler equation} to
\myequation{{\dot v_\gamma} = k({\Theta}+{\Psi}) \,.
\label{eqn:Eulerwithgravity}}
{General relativity says that {$\Phi$} and {$\Psi$} are the relativistic
analogues of the {Newtonian potential} and that {$\Phi \approx - \Psi$} in the
absence of sources of anisotropic stress or viscosity.}

\medskip\noindent\emph{Photon Dominated Oscillator ---}
We can again combine the continuity equation (\ref{eqn:continuitywithgravity})
and Euler equation (\ref{eqn:Eulerwithgravity}) to form the forced simple
harmonic oscillator system
\myequation{{\ddot \Theta} + {c_s^2} k^2 {\Theta} = -\frac{k^2}{3}{\Psi}
 -{\ddot\Phi}\,.
 \label{eqn:oscillatorwithgravity}}
 Note that the effect of baryon inertia is still absent in this system.
 To make further progress in understanding the effect of gravity on acoustic oscillations
 we need to specify the gravitational potential $\Psi \approx -\Phi$ from the Poisson equation
 and understand its time evolution.

\medskip\noindent\emph{Poisson Equation and Constant Potentials ---}
{In our matter-dominated approximation, $\Phi$ is generated by 
 matter density fluctuations $\Delta_m$ through the cosmological
{Poisson equation}}
\myequation{ k^2 {\Phi} = 4\pi G a^2 \rho_m {\Delta_m} \label{eqn:Poisson}\,,}
{where the difference from the usual Poisson equation comes from 
the use of {comoving coordinates} for $k$ ($a^2$ factor), the removal of the
{background density} into the background expansion $(\rho_m  \Delta_m)$ and
finally a {coordinate subtlety} that enters into the definition of $\Delta_m$}.
In general, the relativistic Poisson equation would have contributions from
the momentum density.  Alternately we can relate
\begin{equation}
\Delta_m = {\delta \rho_m \over \rho_m} + 3{\dot a \over a} {v_m \over k}
\end{equation}
to the density fluctuation in a coordinate system
that comoves with the matter (formally through a gauge transformation).
Beyond the matter dominated approximation, the Poisson equation would
carry contributions from all of the species of energy density and the comoving
coordinate system would also reflect the total.

In a matter-dominated epoch (or in fact any epoch when gravitational
potential gradients and not stress gradients dominate the momentum
equation), the matter Euler equation implies
 ${v_m} \sim k\eta {\Psi}$.   The matter continuity equation then implies
${\Delta_m} \sim
- k \eta {v_m} \sim -(k\eta)^2 {\Psi}$.  The Poisson equation then yields ${\Phi }\sim {\Delta_m}/
(k\eta)^2 \sim -{\Psi}$. {Here we have used the {Friedmann equation} $H^2 = 8\pi G\rho_m/3$ and $\eta
= \int d\ln a/(a H) \sim 1/(a H)$.}
  In other words, the density perturbation $\Delta_m$ grows at
exactly the right rate to keep the gravitational potential constant. 
More formally, if {stress perturbations} are negligible compared with {density
perturbations} ( $\delta p \ll \delta \rho$ ) then the potential will remain  constant in
periods where  the background equation of state $p/\rho$ is constant.  
With a varying equation of state, it is the {comoving curvature} $\zeta$ rather than
the Newtonian curvature that is strictly constant  (\citealt{Bar80}).

\medskip\noindent\emph{Effective Temperature ---}
If the gravitational potential is constant, we can rewrite the oscillator
equation (\ref{eqn:oscillatorwithgravity}) as
\myequation{{\ddot \Theta} +{\ddot\Psi} + c_s^2 k^2 ({\Theta} + 
{\Psi}) = 0 \,.}
The solution in Eqn.~(\ref{eqn:oscillatorsolution}) 
for adiabatic initial conditions then generalizes to
\myequation{[{\Theta}+{\Psi}](\eta) = [{\Theta}+{\Psi}](0) \,\cos(k s)\,.
\label{eqn:oscillatorsolutionwithgravity}}
Like a mass on  spring in a constant gravitational field of the earth, the solution just
represents oscillations around a displaced minimum.  

Furthermore,
{$\Theta + \Psi$} is also the {observed temperature fluctuation}.  Photons
		lose energy climbing out of {gravitational potentials} at recombination and
		so the observer at the present will see
		\begin{equation}
		{\Delta T \over T} = \Theta +\Psi \,.
		\end{equation}

Therefore from the perspective of the observer, the acoustic oscillations ignoring
both gravitational forces and gravitational redshifts is unchanged: initial
perturbations in the effective temperature oscillate around zero with a frequency
given by the sound speed.    What the consideration of gravity adds
is
a way of connecting the initial conditions to inflationary curvature fluctuations.

\begin{figure}[tb]
\begin{center}
\includegraphics[width=0.9\textwidth]{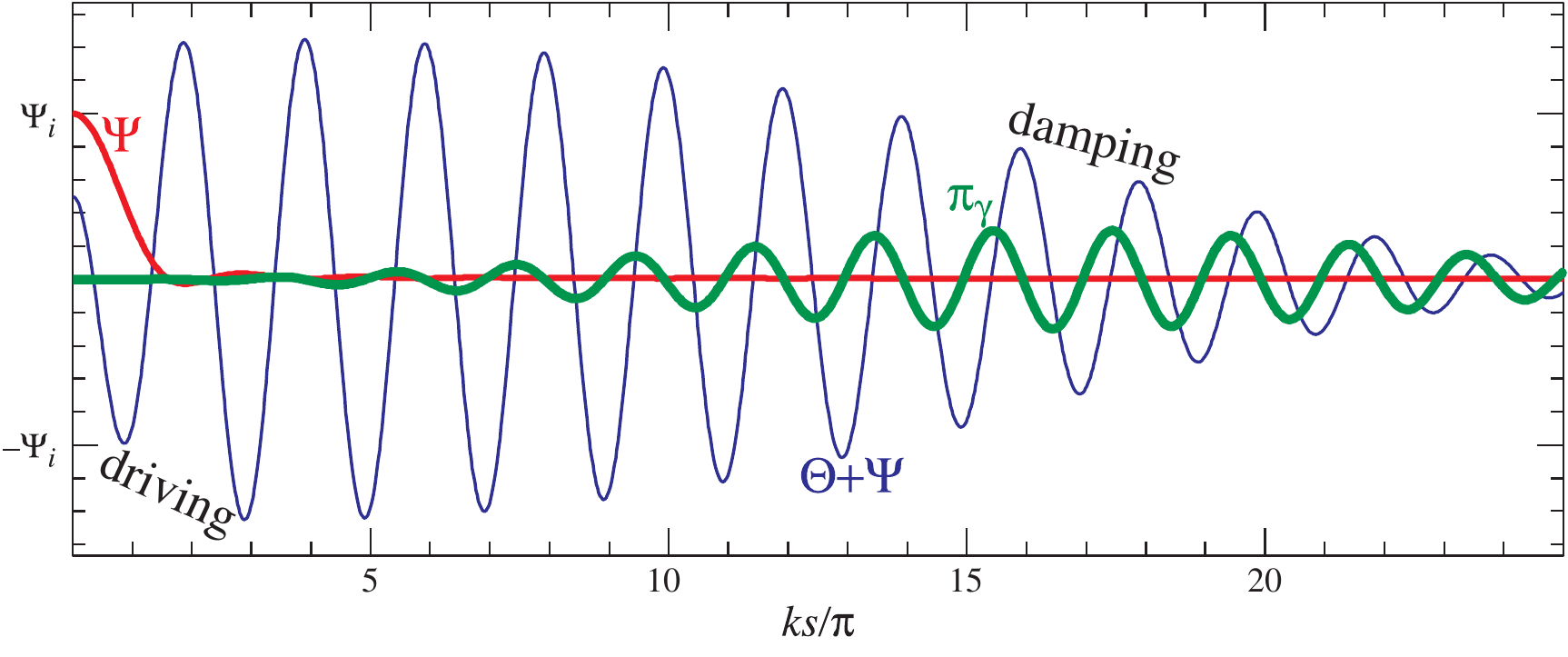}
\caption{Acoustic oscillations with gravitational forcing and dissipational damping.  For
a mode that enters the sound horizon during radiation domination, the gravitational potential
decays at horizon crossing and drives the acoustic amplitude higher.   As the photon 
diffusion length increases and becomes comparable to the wavelength, radiative viscosity
$\pi_\gamma$ is generated from quadrupole anisotropy leading to dissipation and polarization
(see \S\ref{sec:acousticpol}).  Adapted from \cite{HuDod02}.} \label{fig:osc3}
\end{center}
\end{figure}

\medskip\noindent\emph{The Sachs-Wolfe 1/3 ---}  The effective temperature perturbation
includes both the local temperature $\Theta$ in Newtonian coordinates and the gravitational
redshift factor $\Psi$.  For $k s_* \ll 1$, the oscillator is frozen at its initial conditions and
the total is called the \cite{SacWol67} effect.  The division into two pieces is actually 
an artifact of the coordinate system.  Both pieces are determined by the initial curvature
perturbation from inflation. 

To see this, let us relate both the Newtonian potential and local temperature fluctuation
to a change in time coordinate from comoving coordinates to Newtonian coordinates
(formally through a gauge transformation, see \citealt{WhiHu97} for a more detailed treatment).
{The Newtonian {gravitational potential} $\Psi$ is a perturbation to the temporal coordinate 
(see Eqn.~(\ref{eqn:lineelement}))
\myequation{\frac{\delta t}{t} = {\Psi}\,.}
{Given the Friedmann equation, this is equivalent to a perturbation in the
scale factor }
\begin{eqnarray}
{t}  &=& \int {da \over a {H}}  
\propto \int {da \over a {\rho^{1/2}}}  \propto a^{3(1+{w})/2} \,,
\end{eqnarray}
{where {$w \equiv p/\rho$}. During matter domination $w=0$  and}
\myequation{\frac{\delta a}{a} = {\frac{2}{3}}\frac{\delta t}{t}\,.}

{Since the CMB temperature is {cooling} as $T \propto a^{-1}$ a local change in
the scale factor changes the local temperature}
\begin{equation}
\Theta = -{\delta a \over a} = -{2 \over 3}\Psi \,.
\end{equation}
Combining this with $\Psi$ to form the effective temperature gives
\myequation{{\Theta} + {\Psi} = {\frac{1}{3}}{\Psi}\,.}
The consequence is that overdense regions where $\Psi$ is negative  (potential wells)
are cold spots in the effective temperature.  

Inflation provides a source for initial curvature fluctuations.  Specifically the comoving
curvature perturbation $\zeta$ becomes a Newtonian curvature of
\begin{equation}
\Phi = {3 \over 5} \zeta
\end{equation}
in the matter-dominated epoch.   The initial amplitude of scalar curvature perturbations
is usually given as $A_S =\delta_\zeta^{2}$  which characterizes the variance
contribution per e-fold to the curvature near some fiducial wavenumber
$k_n$  (see Eqn.~\ref{eqn:logpower})
\begin{equation}
\Delta_\zeta^2 = {k^3 P_\zeta \over 2\pi^2}= \delta_\zeta^2 \left( {k \over k_n} \right)^{n-1}\,,
\label{eqn:initialpower}
\end{equation}
where $n=1$ for a scale invariant spectrum.  Combining these relations for
$n=1$
\begin{equation}
{\ell (\ell+1) C_\ell \over 2\pi} \approx {\delta_\zeta^2 \over 25} \,.
\label{eqn:COBEnorm}
\end{equation}
in the Sachs-Wolfe 
limit.  The $10^{-5}$ fluctuations measured by COBE then correspond to 
$\delta_{\zeta} \approx 5 \times 10^{-5}$.

\begin{figure}[tb]
\begin{center}
\includegraphics[width=3in]{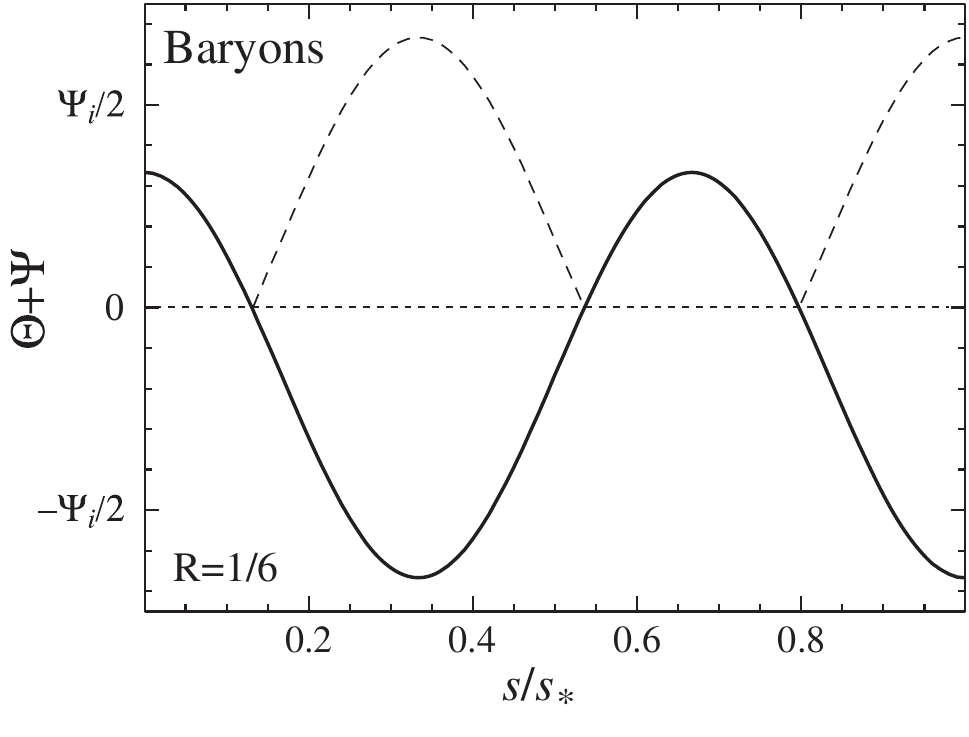}
\caption{Acoustic oscillations with baryons.  Baryons add inertia to the photon-baryon plasma displacing the zero point of the oscillation and making compressional peaks (minima) larger than 
rarefaction peaks (maxima).  The absolute value of the fluctuation in effective
temperature is shown in dotted lines.  Adapted from \cite{HuDod02}.} \label{fig:osc2}
\end{center}
\end{figure}

\subsection{Baryonic Effects}
\label{sec:baryons}

The next level of detail that we need to add is the inertial effect of baryons in the plasma.  
Since equation~(\ref{eqn:Rfactor}) says that the baryon momentum becomes comparable
to the photon momentum near recombination, we can expect order unity corrections
on the basic acoustic oscillation picture from baryons.   With the precise measurements
of the first and second peak, this change in 
the acoustic morphology has already provided the most sensitive measure
of the baryon-photon ratio to date exceeding that of big bang nucleosynthesis.

\medskip\noindent\emph{Baryon Loading ---} Baryons add extra {mass} to the photon-baryon plasma or equivalently an enhancement of the momentum density of the plasma given by
$R=(p_b +\rho_b)/(p_\gamma + \rho_\gamma)$.
{Specifically the momentum density of the {joint system}}
\begin{eqnarray}
(\rho_\gamma  + p_\gamma) {v_\gamma} + (\rho_b + p_b) {v_b}
&\equiv & (1+{R}) (\rho_\gamma + p_\gamma) {v_{\gamma b}} 
\end{eqnarray} 
is conserved.  For generality, we have introduced the baryon velocity $v_b$ and the
momentum-weighted velocity $v_{\gamma b}$ but in the tightly coupled plasma $v_b \approx v_{\gamma b}
\approx  v_\gamma$ (\emph{c.f.} \S\ref{sec:damping}).

The Euler equation (\ref{eqn:Eulerwithgravity}) becomes
\begin{eqnarray}
 [(1+{R})(\rho_\gamma + p_\gamma) {{\bf v}_{\gamma b}} ]^{\cdot} & =&  
 - 4 {\dot a \over a}(1+{R})(\rho_\gamma + p_\gamma){{\bf v}_{\gamma b}} \nonumber\\
&&\qquad -{\nabla p_\gamma} 
 - (1+{R})(\rho_\gamma + p_\gamma){ \nabla \Psi \,.} 
\end{eqnarray}
This equation takes the same form as the photon-dominated system except for the
$(1+{R})$ terms which multiply everything but the pressure gradient terms since the
pressure comes predominantly from the photons.   We can rewrite the equation more
compactly as 
\myequation{ [(1+{R}) {v_{\gamma b}}]^{\cdot} 
= k {\Theta} + (1+{R}) k {\Psi}\,.}

\medskip\noindent\emph{Oscillator with Gravity and Baryons ---}
{The photon continuity} equation (\ref{eqn:continuitywithgravity}) remains the same so
that the oscillator equation becomes 
\myequation{[(1+{R}){\dot \Theta}]^{ \cdot} + {1 \over 3} k^2 {\Theta} 
= -\frac{1}{3} k^2 (1+{R}){\Psi} - [(1+{R}){\dot \Phi}]^\cdot \,.\label{eqn:oscillatorwithbaryons}}
This equation is the final oscillator equation for the tight coupling or perfect fluid regime.

We can make several simplifications to illuminate the impact of baryons.  
First let us continue to use the matter dominated approximation where $\Psi=-\Phi=$const.
Next let us make the adiabatic approximation where the change in $R$ is slow
compared with the frequency of oscillation
 $\dot R/R \ll \omega = k c_s$.  In that case, Eqn.~(\ref{eqn:oscillatorwithbaryons}) looks like an 
 oscillator equation with a fractional change in the mass given by $R$ and a change in the sound
 speed
\begin{equation}
c_s^2 = {\dot p_{\gamma} +\dot p_b  \over \dot \rho_{\gamma} +\dot \rho_b} = {1 \over 3(1+R)} \,.
\end{equation} 
Consequently, the solution in Eqn.~(\ref{eqn:oscillatorsolutionwithgravity}) is modified
as 
\myequation{[{\Theta} + (1+{R}) {\Psi}](\eta) = [{\Theta} + (1+{R}) {\Psi}](0)\cos ks \,.}
This solution is reminiscent of that of adding mass to the spring in a
constant gravitational field of the earth.

There are 3 effects of baryon loading.   First the amplitude of oscillations increases
by a factor of $1+3R$
\begin{equation}
 [{\Theta} + (1+{R}) {\Psi}](0) = {1 \over 3}(1+3R) \Psi(0) \label{eqn:baryonamp}\,.
 \end{equation}
 Next the equilibrium point of the oscillation is now shifted so that relative to zero effective
 temperature, the even and odd peaks have different amplitudes
\begin{eqnarray}
\, [{\Theta} + {\Psi}]_{n} & = &
 \left[  \pm (1+3{R})- 3{R} \right] {1 \over 3}{\Psi(0)} \,, \nonumber\\
\,  [{\Theta} + {\Psi}]_{1} - [{\Theta} + {\Psi}]_{2} &= & 
 [-6 {R}]  {1 \over 3} {\Psi(0)\,.}
\end{eqnarray}
In particular, baryon loading increases the heights of the odd peaks over the even peaks
(see Fig.~\ref{fig:osc2}).

Finally, the lowering of the sound speed changes the acoustic scale in 
Eqn.~(\ref{eqn:thetaA}) as
\myequation{ {\ell_A} \propto \sqrt{1+{R}} \,.}
The effects of baryon loading in a full calculation are actually smaller since
$R$ is growing in time.

\medskip\noindent\emph{Baryon-Photon Momentum Ratio Evolution ---}
One can get a handle on the effect of evolution of $R$ by again equating the
system to the analogous physical oscillator.  The baryons add inertia or mass
to the system and for a slowly varying mass the oscillator equation has an
adiabatic invariant
\myequation{
{E \over \omega} = {1 \over 2} {m_{\rm eff}} \omega A^2 = {1 \over 2} (1+R)
 k {c_s} A^2 \propto A^2 (1+{R})^{1/2} = {\rm const.}}
{Amplitude of oscillation $A \propto (1+{R})^{-1/4}$ {decays adiabatically} as the photon-baryon
ratio changes.  This offsets the gain in the overall amplitude from Eqn.~(\ref{eqn:baryonamp}).
Coupled with uncertainties in the distance to recombination in interpreting the
$\ell_A$ measurement, this leaves the modulation of the peak heights as the
effect that provides most of the information about the baryon-photon ratio in
the acoustic peaks.}

\subsection{Matter-Radiation Ratio}
\label{sec:matter}

Next  we want to go beyond the matter-dominated
expansion approximation.   The universe only becomes matter dominated in the  few e-folds
before recombination (see Eqn.~(\ref{eqn:matterradiation})).  Peaks corresponding to
wavenumbers that began oscillating earlier carry the effects of the prior epoch of
radiation domination.  These effects come in through the evolution of the
gravitational potential which acts as a forcing function on the oscillator through 
Eqn.~(\ref{eqn:oscillatorwithbaryons}).

\medskip\noindent\emph{Potential Decay ---}  The argument given in \S\ref{sec:gravity} for
the constancy of the gravitational potential depends crucially on gravity being the
dominant force affecting the total density.  When radiation dominates the total density,
radiation stresses become more important than gravity on scales smaller than
the sound horizon.   The total density fluctuation stops growing and instead oscillates
with the acoustic frequency.
The Poisson equation in the radiation dominated epoch
\myequation{ k^2 {\Phi} = 4\pi G a^2 \rho_r {\Delta_r}}
then implies that $\Phi$ oscillates and decays with an amplitude $\propto a^{-2}$
(see Fig.~\ref{fig:osc3}).  As an aside, the relativistic stresses of dark energy
make the gravitational potential decay again during the acceleration epoch and 
lead to the so called integrated Sachs-Wolfe effect.

\medskip\noindent\emph{Radiation Driving ---} An examination of Fig.~\ref{fig:osc3} shows
that the time evolution of the gravitational potential is in phase with the acoustic
oscillations themselves and act a driving force on the acoustic oscillations.  We
can estimate the effect on the amplitude of oscillations in the limit that the force
is fully coherent.  In that case we can take the continuity equation (\ref{eqn:continuitywithgravity})
and simply integrate it 
\begin{eqnarray}
[{\Theta} +{\Psi}](\eta) & = [{\Theta} + {\Psi}](0) + { \Delta \Psi} - 
{\Delta \Phi} \nonumber \\
			        & = {1 \over 3} {\Psi(0)} -2 {\Psi(0)}  = {5 \over 3}{\Psi(0)} \,.
\end{eqnarray}
This estimate gives an acoustic amplitude that is 
{{5$\times$} that of
		the Sachs-Wolfe effect.}
This enhancement only occurs for modes that begin oscillating during the
radiation dominated epoch, \emph{i.e.}~the higher peaks.  The net effect is a gradual
ramp up of the acoustic oscillation amplitude across the horizon wavenumber
at matter radiation equality (see Fig.~\ref{fig:dop}).

In fact the coherent
{approximation is {exact} for a photon-baryon fluid but must be corrected
for the neutrino contribution to the radiation density.

\medskip\noindent\emph{External Potential Approach ---}  For pedagogical purposes
it is sometimes useful to go beyond the coherent approximation.   Neutrino corrections
are one example; isocurvature initial conditions are another.

{We have seen that the solutions to {homogeneous equation} for the oscillator
equation (\ref{eqn:oscillatorwithgravity}) are}
\myequation{(1+R)^{-1/4} { \cos(ks})\,, \qquad
                  (1+R)^{-1/4} {\sin(ks)} } 
                  in the adiabatic or high frequency limit.
{Considering the potentials as external, we can solve for the
temperature perturbation as (\citealt{HuSug95a})}
\begin{eqnarray}
(1+R)^{1/4} {\Theta(\eta)} &= 
\Theta(0) {\cos(ks)} +
{\sqrt{3} \over k} \left[ \dot\Theta(0) + {1 \over 4} \dot R(0) \Theta(0) \right] {\sin k s}\nonumber \\
&\qquad + {\sqrt{3} \over k} \int_0^\eta d\eta' (1+R')^{3/4} {\sin[k s - ks']}  {F(\eta')\,,} 
\end{eqnarray}
{where}
\myequation{ {F} = - {\ddot \Phi} - {\dot R \over 1+R} {\dot \Phi} - {k^2 \over 3} {\Psi}\,. }
By including the neutrino effects in the gravitational potential, we can show from this
approach that radiation driving actually creates an acoustic amplitude that is
close to $4\times$ the Sachs-Wolfe effect.

\subsection{Damping}
\label{sec:damping}

The final piece in the acoustic oscillation puzzle is the damping of power beyond 
$\ell \sim 10^{3}$ shown in Fig.~\ref{fig:filter}.  Up until this point, we have considered
the oscillations in the tight coupling approximation where the photons and baryons respond to pressure and gravity as a single perfect fluid.  
Fluid imperfections are associated with the Compton mean free path in Eqn.~(\ref{eqn:mfp}).
Dissipation becomes strong at the diffusion scale, the distance a photon can random
walk in a given time $\eta$ (\citealt{Sil68})
\myequation{ {\lambda_D} = \sqrt{N} { \lambda_C }= \sqrt{\eta/{\lambda_C}}
		\,{ \lambda_C}
			= \sqrt{\eta{\lambda_C}}\,.}
{This scale is the {geometric mean} between the horizon and mean free path.  Given that
 $\lambda_D /\eta_* \sim$ { few percent}, we expect that the {$n\ge $ 3} peaks to 
		be affected by {dissipation}.}  To improve on this estimate, we develop
		the microphysical description of dissipation next.

\medskip\noindent\emph{Continuity Equations ---} To treat the photons and baryons 
as separate systems, we now need to supplement the photon continuity equation with
the baryon continuity equation
\myequation{
{\dot \Theta} = -\frac{k}{3} {v_\gamma} - {\dot \Phi}\,, \quad
{\dot \delta_b} = -k{ v_b} - 3 {\dot \Phi}\,. }
The baryon equation follows from  
	 	number conservation with $\rho_b = m_b n_b$ and $\delta_{b}\equiv \delta\rho_{b}/\rho_{b}$.}

\medskip\noindent\emph{Euler and Navier-Stokes Equations ---} The momentum conservation
equations must also be separated into photon and baryon pieces
\eqnsize\begin{eqnarray*}
{\dot v_\gamma}  &=& k({\Theta} + {\Psi}) - \frac{k}{6}{\pi_\gamma} - \dot \tau ({v_\gamma} - {v_b})\,, \\
{\dot v_b}   &=& - \frac{\dot a }{a} {v_b} + k{\Psi} + \dot \tau({v_\gamma} -{ v_b})/R \,,
\label{eqn:jointcontinuity}
\end{eqnarray*}
{where the photons gain an anisotropic stress term ${\pi_\gamma}$ from
{radiation viscosity}.   The baryon equation follows from the same derivation as in \S\ref{sec:basics} where the redshift of the momentum is carried by the bulk velocity instead
of the redshifting temperature.
 Finally there is a {momentum exchange} term from Compton 
scattering.  Note that the total momentum in the system is
conserved and hence the scattering terms come with opposite sign.

\medskip\noindent\emph{Viscosity ---}  Radiative shear viscosity is equivalent to quadrupole
moments in the temperature field.  These quadrupole moments are generated by 
radiation streaming from hot to cold regions much like how temperature inhomogeneity
are converted to anisotropy in Fig.~\ref{fig:projection}.

In the tight coupling limit where $\dot\tau/k$, the optical depth through a wavelength of
the fluctuation is high 
one therefore expects
\myequation{{\pi_\gamma} \sim {v_{\gamma}} { k \over \dot \tau}\,,}
{since it must be generated by streaming and suppressed by scattering.
A more detailed calculation from the Boltzmann or {radiative transfer} equation says
(\citealt{Kai83})}
\myequation{{\pi_\gamma} \approx 2 A_v {v_{\gamma}} { k \over \dot \tau}\,,}
{where $A_v=16/15$ once polarization effects are incorporated}
\myequation{
{\dot v_\gamma}  = k({\Theta} + {\Psi}) - \frac{k}{3} A_v {\frac{k}{\dot\tau}} {v_\gamma}\,. }
The oscillator equation with viscosity becomes
\begin{eqnarray*}
 {c_s^2} \frac{d}{d\eta} ( {c_s^{-2}} {\dot \Theta})
		+{\frac{k^2 c_s^2}{\dot \tau} }A_v{\dot \Theta}+ k^2 {c_s^2} {\Theta} =
		-\frac{k^2}{3} {\Psi} -
		{c_s^2} \frac{d}{d\eta} ( {c_s^{-2}} {\dot \Phi}) \,,
\end{eqnarray*}
As in a mechanical oscillator, a term that depends on $\dot\Theta$ provides a dissipational
term to the solutions.

\medskip\noindent\emph{Heat Conduction ---} Relative motion between the photons
and baryons also damp oscillations.  By expanding the continuity and momentum conservation
equations in the small number $k/\dot\tau$ one obtains for the full oscillator equation
\begin{eqnarray*}
 {c_s^2} \frac{d}{d\eta} ( {c_s^{-2}} {\dot \Theta})
		+{\frac{k^2 c_s^2}{\dot \tau}[A_v + A_h] }{\dot \Theta}+ k^2{c_s^2} {\Theta} =
		-\frac{k^2}{3} {\Psi} -
		{c_s^2} \frac{d}{d\eta} ( {c_s^{-2}} {\dot \Phi})\,
		\end{eqnarray*}
where
		\myequation{A_h = {R^2 \over 1+R}\,.}

\medskip\noindent\emph{Dispersion Relation ---}  We can solve the damped oscillator
equation in the adiabatic approximation by taking a trial solution $\Theta 
\propto \exp(i \int \omega d\eta)$ to obtain the dispersion relation
\begin{eqnarray}
\omega	    = \pm k c_s \left[ 1 \pm  {i \over 2} {k c_s \over \dot \tau}(A_v + A_h) \right]  \,.
\end{eqnarray}
The imaginary term in the dispersion relation gives an exponential damping of the 
oscillation amplitude
\begin{eqnarray}
 \exp(i \int {\omega} d\eta) &=& e^{\pm i k s} 
\exp[-k^2 \int d\eta {1 \over 2}{c_s^2 \over \dot\tau} (A_v+A_h) ] \nonumber \\
&=& e^{\pm i k s} \exp[-(k/{k_D})^2] \,,
\end{eqnarray}
where the 
diffusion wavenumber is given by
\begin{equation}
{k_D^{-2}} = \int { d\eta { 1 \over \dot \tau}}{1 \over 6(1+R)} \left( {16 \over 15} +
	{R^2 \over (1+R) } \right)\,.
	\end{equation}
{Note that in both the high and low $R$ limits}
\begin{eqnarray}
\lim_{R \rightarrow 0} {k_D^{-2}} &=& {1\over 6}  {16 \over 15}
\int {{d\eta} {1 \over \dot \tau}} \,, \nonumber \\ 
\lim_{R \rightarrow \infty} {k_D^{-2}} &=& {1 \over 6}\int {{d\eta} {1 \over \dot \tau} \,.} 
\end{eqnarray}
{Hence the dissipation scale is 
\myequation{{\lambda_D} = {2\pi \over {k_D}}
 \sim {2 \pi \over \sqrt{6} }
({\eta \dot\tau^{-1}})^{1/2} }
and comparable to the geometric mean between horizon and mean free path as expected from the {random
walk} argument\,.}  For a baryon density of $\Omega_{b}h^{2} \approx 0.02$, radiation
viscosity is responsible for most of the dissipation and we show the correspondence
between viscosity generation and dissipation in Fig.~\ref{fig:osc3}.

Since the diffusion length changes rapidly through recombination and the medium
changes from optically thick to optically thin, the damping estimates above are only 
qualitative.   A full Boltzmann (radiative transfer solution) shows a more gradual, but
still exponential, damping of roughly 
\begin{equation}
{\cal D}_{\ell} \approx \exp[ -(\ell/\ell_{D})^{1.25} ]\,,
\end{equation}
with a damping scale of $\ell_D = 2\pi D_*/\lambda_D$ and (\citealt{Hu04b})
\begin{eqnarray}
{\lambda_D \over {\rm Mpc} }\approx 64.5 
\left( {\Omega_m h^2 \over 0.14} \right)^{-0.278} \left( {\Omega_b h^2 \over 0.024} \right)^{-0.18}\,,
\end{eqnarray} 
for small changes around the central values of $\Omega_m h^2$ and $\Omega_b h^2$.

This envelope also accounts for enhanced damping due to the finite duration of recombination.
Instead of a delta function in the projection equation (\ref{eqn:thetadelta}) we have
the visibility function $\dot \tau e^{-\tau}$ that acts as a smearing out of any 
contributions with
wavelengths shorter than the thickness of the recombination surface that survive dissipation.

\subsection{Information from the Peaks}
\label{sec:information}

In the preceding sections we have examined the physical processes involved in
the formation of the acoustic peaks and explained their sensitivity
to the energy content and expansion rate of the universe.   Converting the measurements
into parameter constraints of course requires a more accurate numerical description.
Numerical codes that solve the Einstein-Boltzmann radiative transfer equations for
the CMB and matter (\citealt{PeeYu70, BonEfs84, VitSil84}) are now accurate at the
$\sim 1\%$ level for the acoustic peaks for publically available codes
(\citealt{SelZal96,Lewetal00}).   Their numerical precision on the other
hand is substantially better and approaches the $0.1\%$ level required for cosmic variance
limited measurements out to $\ell \sim 10^3$. The accuracy is  now limited by the input physics, 
mainly
recombination (see \ref{sec:recombination}).  
In this section, we relate the qualitative discussion of
the previous sections to the quantitative information content of the peaks.

\begin{figure}[tb]
\begin{center}
\includegraphics[width=3.0in]{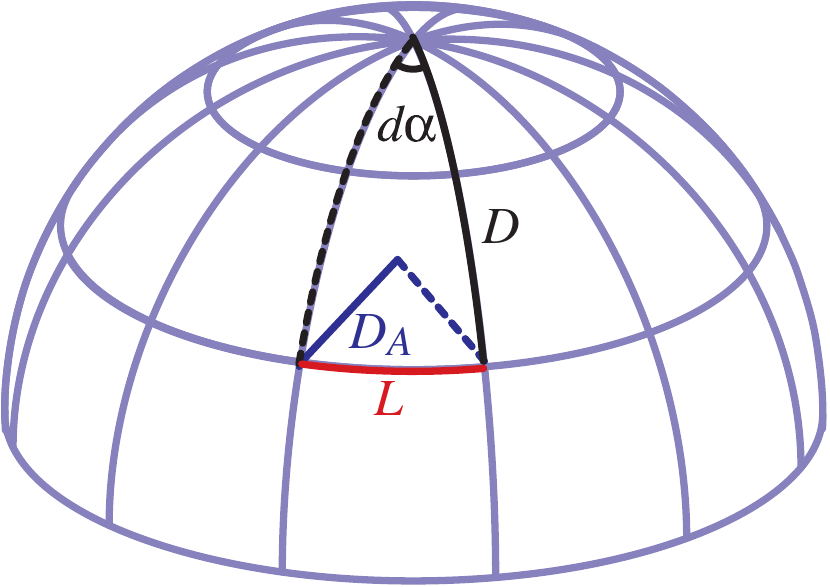}
\caption{Angular diameter distance and curvature.   In a non-flat (here closed) universe, the
apparent or angular diameter distance $D_A = L d\alpha$  does not equal the radial distance traveled by the photon. Objects in a closed universe are further than they appear whereas
in an open universe they are closer than they appear.} \label{fig:da}
\end{center}
\end{figure}

\medskip\noindent\emph{First Peak: Curvature and Dark Energy ---}
The comparison between the predicted acoustic peak scale $\lambda_A$ and its
angular extent provides a measurement of the angular diameter distance
to recombination.   The angular diameter distance in turn depends on the
spatial curvature and expansion history of the universe.  

Sensitivity to the expansion history during the acceleration epoch comes about through
the radial distance a photon travels along the line of sight
\begin{equation}
D_* = \int_0^{z_*} {dz \over H(z)}	= \eta_0-\eta_*	\,.
\end{equation}
With the matter and radiation energy densities measured, the remaining
contributor to the expansion rate $H(z)$ is the dark energy.

However
{in a {curved universe}, the apparent or {angular diameter distance} $D_{A}$ in
Eqn.~(\ref{eqn:thetaA})
		is no longer the distance a photon travels radially along the line of
		sight.
		The radius of curvature of the space is given in terms of the
		total density $\Omega_{\rm tot}$ in units of the critical density as
\begin{equation}	
		 R^{-2} = H_0^2 (\Omega_{\rm tot}- 1) \,.
\end{equation}
		A positive curvature space has $\Omega_{\rm tot}>1$ and a real radius
		of curvature.  A negatively curved space has $\Omega_{\rm tot}<1$ and
		an imaginary radius of curvature.
		The positively curved space is shown in Fig.~\ref{fig:da}.   The curvature makes
		a transverse distance $L$ related to its angular extent $d\alpha$
		as $L = d\alpha D_{A}$ with
\begin{equation} 
D_A = R \sin (D /R) \,.
\end{equation}
The same formula applies for negatively curved spaces but is more conveniently expressed
with the relation  
\begin{equation}
R\sin(D/R) = |R| \sinh (D/|R|)
\end{equation}
for imaginary $R$.
{In a positively curved geometry $D_A < D$ and objects are further than they appear.}
{In a negatively curved universe $R$ is imaginary and $R\sin(D/R) = 
i |R| \sin(D/i|R|) = |R| \sinh (D/|R|)$ -- and $D_A > D$ objects are closer than they appear.}
Since the detection of the first acoustic peak it has been clear that the universe
is close to spatially flat (\citealt{Miletal99,deBetal00,Hanetal00}).  How close and how well-measured $D_*$ is
for dark energy studies depends on the calibration of the physical scale $\lambda_A=2 s_*$,
i.e. the sound horizon at recombination.

\begin{figure}[tb]
\begin{center}
\includegraphics[width=\textwidth]{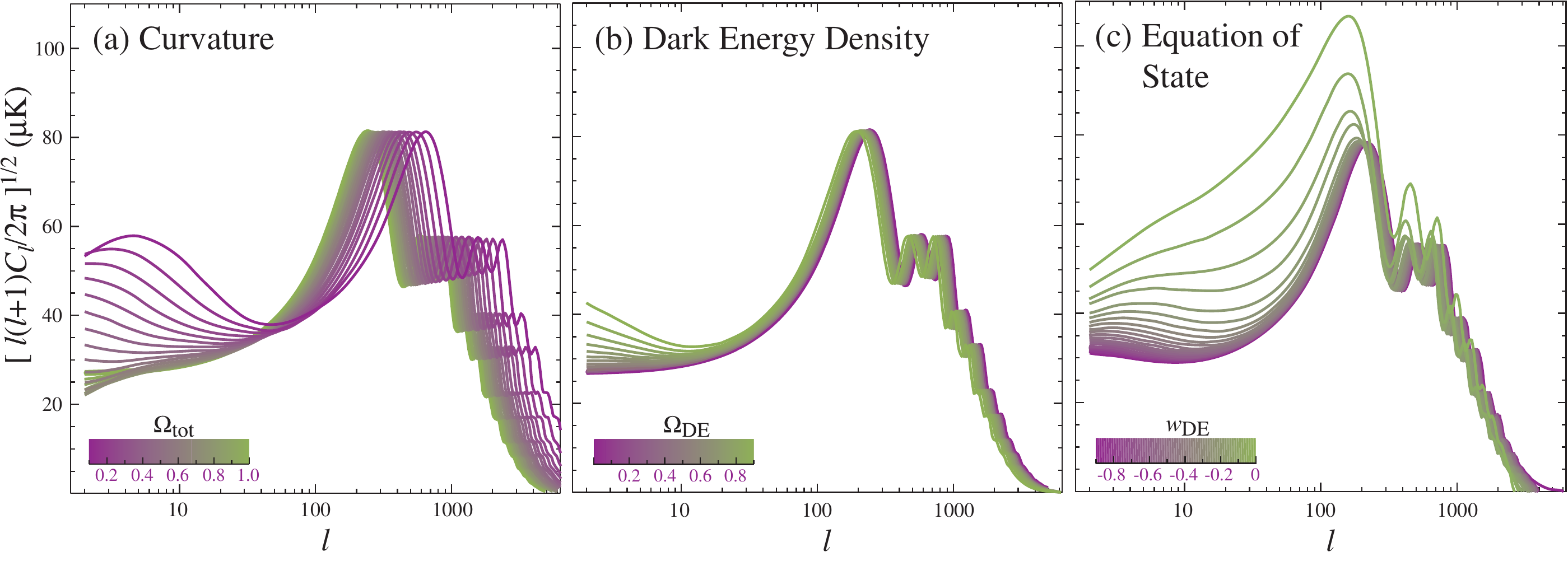}
\caption{Curvature and dark energy.  Given a fixed physical scale for the acoustic peaks
(fixed $\Omega_b h^2$ and $\Omega_m h^2$) the observed angular position of the
peaks provides a measure of the angular diameter distance and the parameters it depends on:
curvature, dark energy density and dark energy equation of state.  Changes at low $\ell$ multipoles
are due to the decay of the gravitational potential after matter domination from the integrated
Sachs-Wolfe effect.} \label{fig:curvdark}
\end{center}
\end{figure}

The sound horizon in turn depends on two things,
\begin{equation}
s_*        = {2 \sqrt{3} \over 3} \sqrt{ a_* \over R_* \Omega_m H_0^2}  \ln
        {\sqrt{1+R_*} +  \sqrt{R_* + r_* R_*}
        \over  1 + \sqrt{r_* R_*}}  \,,
\end{equation} 
 the baryon-photon momentum density ratio
\begin{equation}
R_{*} \equiv {3 \over 4}{\rho_b \over \rho_\gamma} \Big|_{a_*}=  0.729 \left( {  \Omega_{b} h^{2} \over 0.024 } \right) 
\left( {a_* \over 10^{-3} }\right)  \,,
\end{equation}
 and the expansion rate prior to recombination which is determined by the matter radiation ratio
 \begin{equation}
r_{*} \equiv {\rho_r \over \rho_m} \Big|_{a_*}= 0.297 \left( {\Omega_{m}h^{2} \over 0.14 } \right)^{-1} 
\left( {a_{*} \over 10^{-3} }\right)^{-1} \,.
\end{equation}
 The calibration of these two quantities involves the higher acoustic peaks.   The bottom line
 is that the limiting factor in the calibration is the precision with which the matter density is known
 \begin{equation}
{\delta  D_{A*} \over D_{A*} }\approx {1\over 4} {\delta (\Omega_{m}h^{2})\over \Omega_{m}h^{2} }
\label{eqn:distancematter} \,.
\end{equation}
 In principle the Planck satellite can achieve a 1\% measurement of the matter density or
 a $\sim 0.25\%$ measure of distance.  
Current errors on the distance are $\sim $ 1-2\%.

\begin{figure}[tb]
\begin{center}
\includegraphics[width=5in]{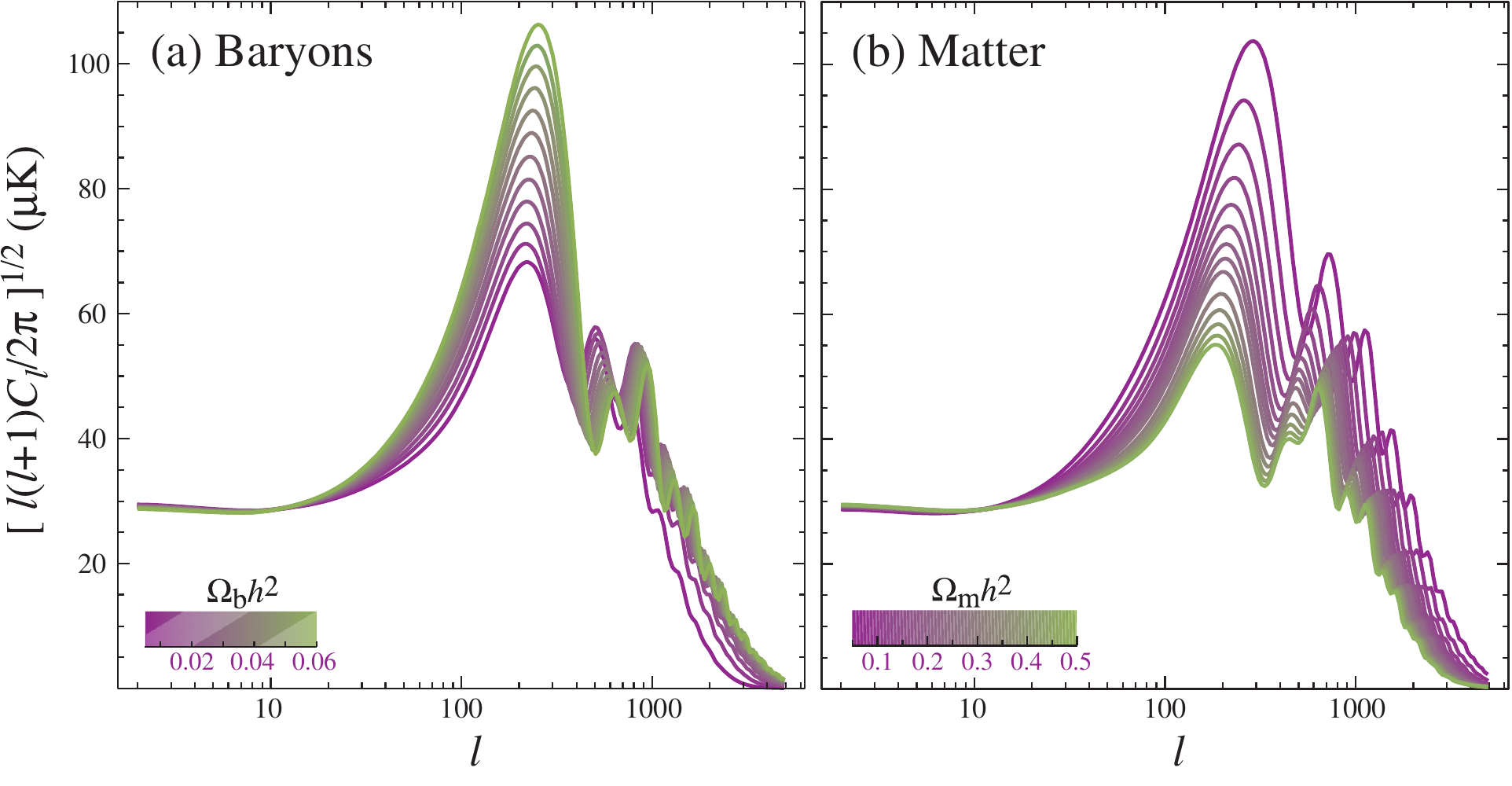}
\caption{Baryons and matter.  Baryons change the relative heights of the even and odd peaks through
their inertia in the plasma.   The matter-radiation ratio also changes the overall amplitude of
the oscillations from driving effects.  Adapted from \cite{HuDod02}.} \label{fig:matterbaryons}
\end{center}
\end{figure}

\medskip\noindent\emph{Second Peak: Baryons ---} The baryon-photon ratio
controls the even-odd modulation of peak heights through the baryon loading
effect (see \S\ref{sec:baryons}).  The second peak represents rarefaction of the
acoustic wave in a gravitational potential and hence is suppressed in amplitude by the
baryon inertia.  The dependence of the spectrum on the baryon density $\Omega_b h^2$
is shown in Fig.~\ref{fig:matterbaryons}.  Since the first tentative detection  of the second
peak (\citealt{deBetal00}) the CMB has placed limits on the baryon density.  
Currently its measurement, largely from WMAP (\citealt{Speetal06}), gives $\Omega_b h^2 = 0.0227 \pm 0.0006$ (\citealt{Reietal08}).  This is well enough
constrained that associated errors on the sound horizon and shape of the matter power
spectrum (see below) are small.

\medskip\noindent\emph{Third Peak: Dark Matter ---}
The third peak begins to show the effects of the matter-radiation ratio on the overall
amplitude of the acoustic peaks.   Furthermore, decay in the gravitational potential
during radiation domination would reduce the baryon loading effect and change
the peak height ratios of the second and third peaks (\emph{e.g.} \citealt{HuFukZalTeg01}).
The dependence of the spectrum on the baryon density $\Omega_m h^2$
is shown in Fig.~\ref{fig:matterbaryons}.
Constraints on the third peak from the DASI experiment 
(\citealt{Pryetal01}) represented the first direct evidence for dark matter at the epoch of
recombination.  Current constraints from a combination of WMAP and higher resolution
ground and balloon based data yield $\Omega_m h^2 = 0.135\pm 0.007$ (\cite{Reietal08}).
Since this parameter controls the error on the distance to recombination through equation
(\ref{eqn:distancematter}) and the matter power spectrum (see below), it is important to improve the precision of its measurement
with the third  higher peaks.

\medskip\noindent\emph{Damping Tail: Consistency---} Under the standard thermal history
of \S\ref{sec:thermal} and matter content, 
the parameters that control the first 3 peaks also determine the
structure of the damping tail at $\ell > 10^3$: namely, the angular diameter distance
to recombination $D_*$, the baryon density $\Omega_b h^2$ and the matter density $\Omega_m h^2$.    When the damping tail was first discovered by the CBI experiment 
(\citealt{Padetal01}), it
supplied compelling support for the standard theoretical modeling of the physics
at recombination outlined here.   Currently the best constraints on the damping tail
are from the ACBAR experiment (\citealt{Reietal08}, see Fig.~\ref{fig:cldata}).  Consistency
between the low order peaks and the damping tail can be used to make precision tests
of recombination and any physics beyond the standard model
at that epoch.  For example, damping tail measurements
can be used to constrain the evolution of the fine structure constant.

\begin{figure}[tb]
\begin{center}
\includegraphics[width=4in]{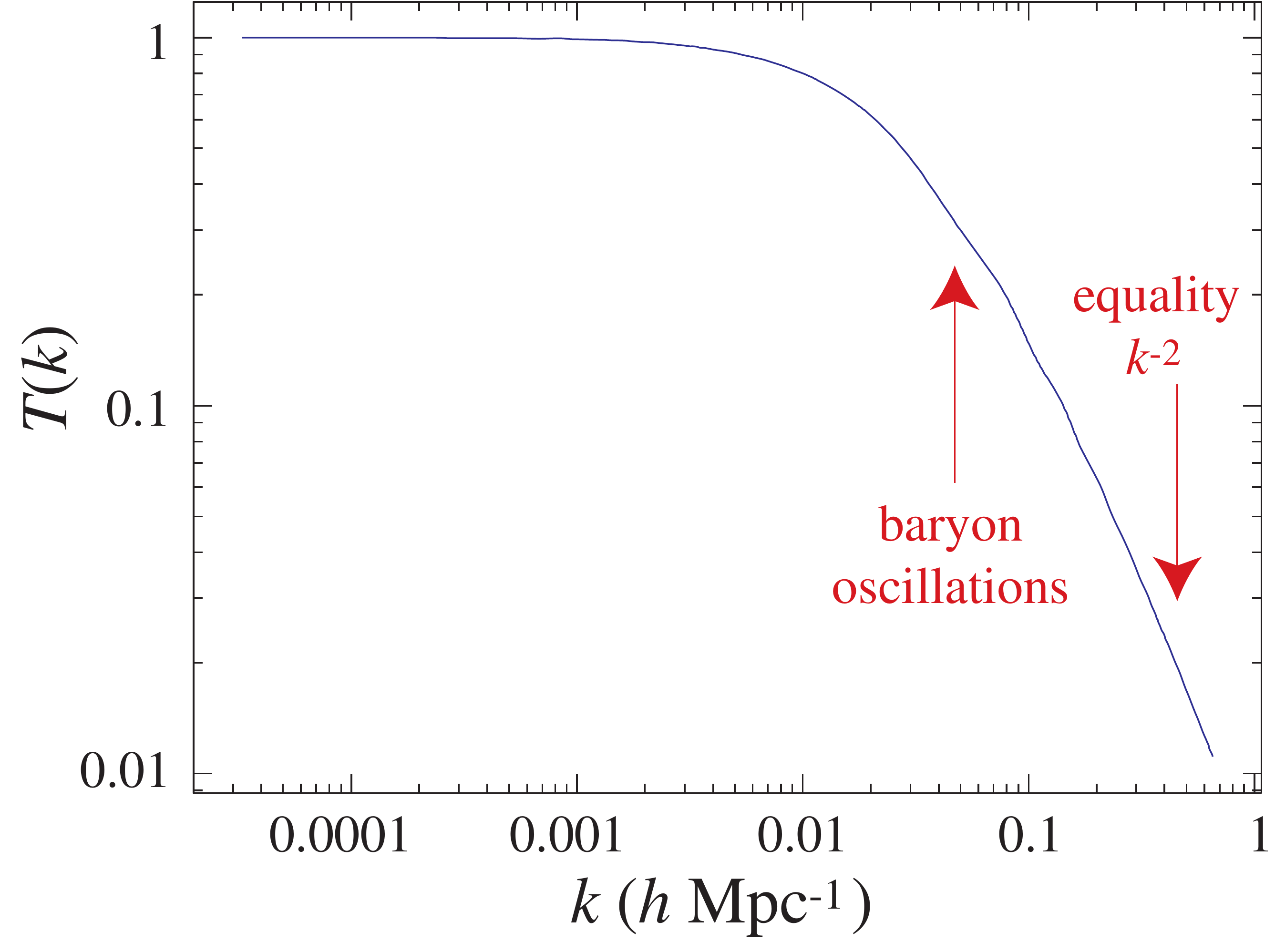}
\caption{Transfer function.  Acoustic and radiation physics is imprinted on the matter
power spectrum as quantified by the transfer function $T(k)$.  The former is responsible
for baryon oscillations in the spectrum and the latter a suppression of growth due to Jeans
stability for scales smaller than the horizon at matter-radiation equality.}\label{fig:tk}
\end{center}
\end{figure}

\medskip\noindent\emph{Matter Power Spectrum: Shape \& Amplitude ---}  The acoustic
peaks also determine the shape and amplitude of the matter power spectrum.  Firstly, acoustic
oscillations are shared by the baryons.  In particular, the plasma motion kinematically produces
enhancements of density near recombination (see Eqn.~\ref{eqn:jointcontinuity}))
\begin{equation}
\delta_b \approx -k\eta_* v_b(\eta_*) \approx -k\eta_*  v_\gamma(\eta_*) \,.
\end{equation}
This enhancement then imprints into the matter power spectrum at an amplitude
reduced by $\rho_b/\rho_m$
due to the small baryon fraction (\citealt{HuSug96}).   Secondly, the gravitational
potentials that the cold dark matter perturbations fall in are 
 evolving through the plasma epoch due to the processes described in \S\ref{sec:matter}.   
 On scales above
the horizon, relativistic stresses are never important and
the gravitational potential remains constant.  The Poisson equation then implies that
\begin{equation}
\Delta \sim (k\eta)^2 \Phi(0) \,,
\end{equation}
and in particular, the density perturbation at horizon crossing where $k\eta \sim 1$ is
$\Delta = \Delta_H \approx \Phi(0)$.
For a fluctuation that crosses the horizon during radiation domination, the total density perturbation
is Jeans stabilized until matter radiation equality 
\begin{equation}
\eta_{\rm eq} \approx 114 \left({\Omega_m h^2 \over 0.14}\right)^{-1} {\rm Mpc} \,.
\end{equation}
Thereafter, relativistic
stresses again become irrelevant and the potential remains unchanged until matter
ceases to dominate the expansion
\begin{equation}
\Phi \approx (k\eta_{\rm eq})^{-2} \Delta_H \sim (k\eta_{\rm eq})^{-2} \Phi(0) \,.
\end{equation}
The transfer in shape from the initial conditions due to baryon oscillations and matter
radiation equality is usually encapsulated into
a transfer function $T(k)$.  Given an initial power spectrum of the form (\ref{eqn:initialpower}),
the evolution through to matter domination transforms the potential power spectrum to
 $k^3 P_\Phi/2\pi^2 \propto k^{n-1} T^2(k)$ where $T(k) \propto k^{-2}$ beyond the 
 wavenumber at matter-radiation equality.   This scaling is slightly modified due to the logarithmic
 growth of dark matter fluctuations during the radiation epoch when the radiation density is
 Jeans stable.
The matter power spectrum and potential power spectrum are related by the
Poisson equation and so carry the same shape.   Specifically
\begin{eqnarray}
 {k^3P_m(k,a) \over 2\pi^2}  
	    &=& {4 \over 25} \delta_\zeta^2 \left({G(a) a \over \Omega_m}\right)^2 { \left( k \over H_0 \right) }^{4}\left( {k \over k_{\rm
norm}}\right)^{n-1} T^2(k)\,,
\end{eqnarray}
where we have included a factor $G(a)$ to account for the decay in the potential during the
acceleration epoch when relativistic stresses are again important (see e.g. \cite{Hu04b}).  This factor only depends on
time and not scale as long as the scale in question is within the Jeans scale of the
accelerating component. In this limit, $G(a)$ is determined by the solution to
\begin{eqnarray}
\frac{d^2 G}{d\ln a^2} + \left( 4 + {d \ln H \over d\ln a} \right) \frac{d G}{d\ln a} 
+ \left[ 3+ {d \ln H \over d\ln a}  - {3\over 2}\Omega_m(a) \right] G = 0 \,,
\end{eqnarray}
with an initial conditions of $G(\ln a_{\rm md})=1$ and $G'(\ln a_{\rm md})=0$ at an epoch
$a_{\rm md}$ when the universe is fully matter dominated.

The transfer function $T(k)$, with $k$ in Mpc$^{-1}$, depends only on the baryon density $\Omega_bh^2$ and the
matter density $\Omega_mh^2$ which are well determined by the 
CMB acoustic peaks.   Features in the matter power spectrum, especially the baryon 
oscillations, then serve as standard rulers for distance measurements \cite{EisHuTeg99a}.
For example, its measurement in a local redshift survey where the distance is calibrated
in $h$ Mpc$^{-1}$ would give the Hubble constant $h$.  Detection of the features
requires a Gpc$^3$ of volume and so precise, purely local, measurements are not feasible.
Nonetheless, 
the first detection of these oscillations by the SDSS LRG redshift survey out to $z\sim 0.4$ 
provide remarkably tight constraints on the acceleration of the expansion (\citealt{Eisetal05}).

The CMB also determines the initial normalization $\delta_\zeta$ and so provides a means
by which to test the effect of the acceleration on the growth function $G(a)$.  The precision of this
determination is largely set by reionization.  The opacity provided by electrons after reionization
suppress the observed amplitude of the peaks relative to the initial amplitude and hence 
\begin{equation}
\delta_\zeta \approx 4.6  e^{-(0.1-\tau)} \times 10^{-5}\,,
\end{equation}
where $\tau$ is the optical depth to recombination.  Note that this is the normalization at
$k =0.05$ Mpc$^{-1}$ and even with uncertainties in the optical depth of $\delta \tau \sim 0.03$
it exceeds the precision of  the COBE normalization
(\emph{c.f.}~Eqn.~(\ref{eqn:COBEnorm})).
Finally, with the matter and baryon transfer effects determined, the acoustic spectrum
also constrains the tilt.   WMAP provided the first hints of a small deviation from
scale invariance (\citealt{Speetal06}) and the current constraints are $n \approx 0.965 \pm 0.015$.

Combining these factors into the conventional measure of the amplitude of
matter fluctuations today
\begin{eqnarray}
\sigma_8^2 &\equiv& \int {dk \over k} {k^{3}P(k,a=1) \over 2\pi^{2}}W_\sigma^2(kr)\nonumber\\
\sigma_{8} &\approx& {\delta_{\zeta} \over 5.59\times10^{-5}} 
\left( { \Omega_{b}h^{2} \over 0.024} \right)^{-1/3}
\left( { \Omega_{m}h^{2} \over 0.14} \right)^{0.563}\nonumber\\
&&\times(3.123h)^{(n-1)/2} \left( { h \over 0.72} \right)^{0.693}  
{G_0 \over 0.76}\,,
\label{eqn:sigma8}
\end{eqnarray}
where $W_\sigma(x)= 3x^{-3}(\sin x - x \cos x)$ is the Fourier transform of
a top hat window of radius $r=8 h^{-1}$Mpc.


\begin{figure}[tb]
\begin{center}
\includegraphics[width=4in]{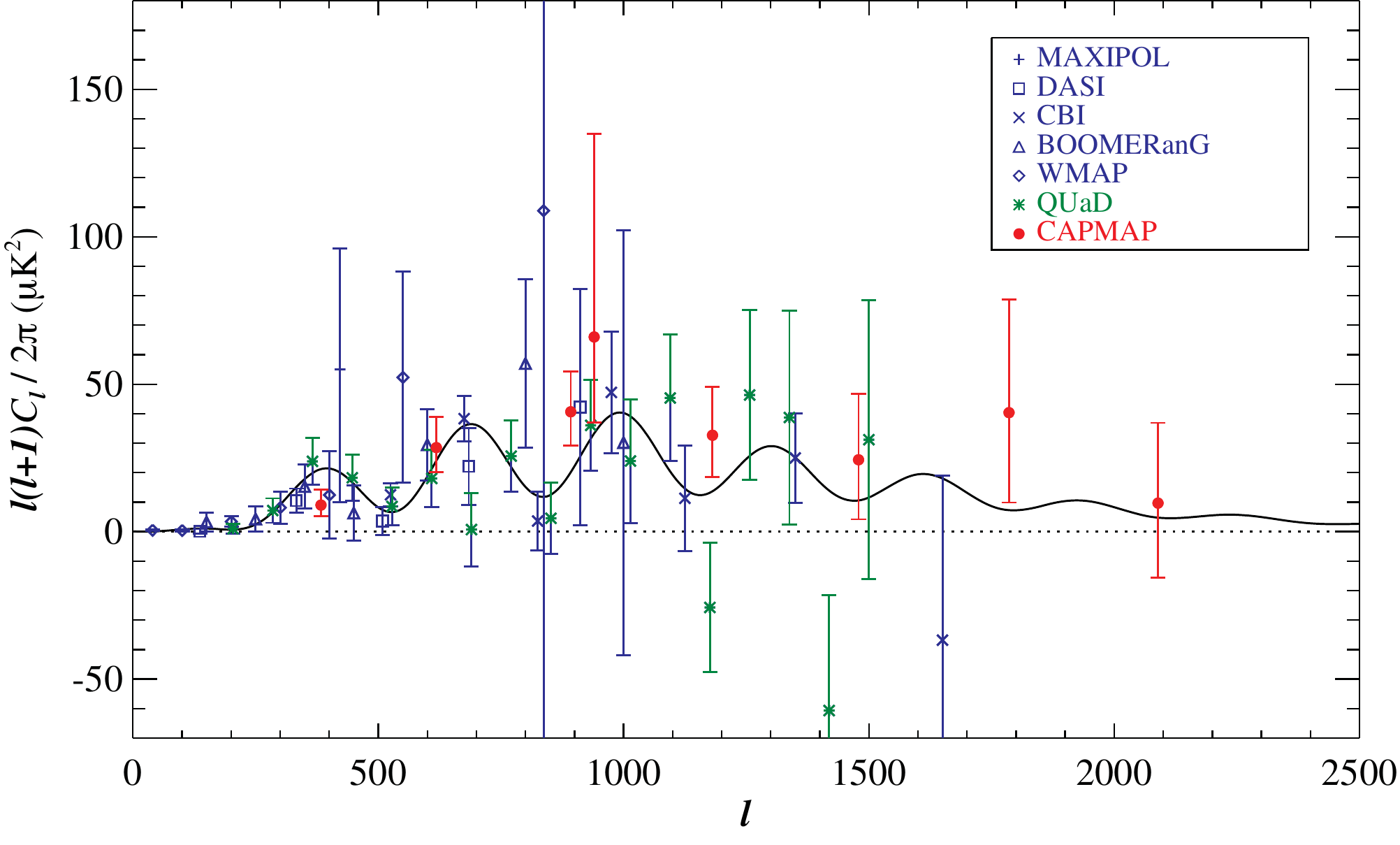}
\caption{$E$-mode polarization power spectrum measurements. Adapted from \cite{Bisetal08}}. \label{fig:clee}
\end{center}
\end{figure}

\section{Polarization Anisotropy from Recombination}
\label{sec:polarization}

Thomson scattering of quadrupolarly anisotropic but unpolarized radiation generates linear polarization.
As we have seen in \S\ref{sec:projection}, $\ell \ge 2$ anisotropy develops only in optically
thin conditions.  Given that polarization also requires scattering to be generated, the polarization
anisotropy is generically much smaller than the temperature anisotropy.   The main source
of polarization from recombination is associated with the acoustic peaks in temperature.
This source was first detected  by the DASI experiment (\citealt{Kovetal02}) and recent years
have seen increasingly precise measurements 
(see Fig.~\ref{fig:clee}).  

Since acoustic polarization arises from linear scalar perturbations,
they possess a symmetry that relates the direction of polarization to the wavevector
or change in the polarization amplitude.   As wee shall see in the next section, 
this symmetry is manifest in the the absence of $B$-modes
(\citealt{KamKosSte97, ZalSel97}).
$B$-modes at recombination can be generated from the quadrupole moment of a gravitational
wave.  This yet-to-be-detected signal would be invaluable for early universe studies involving
the inflationary origin of perturbations.

We begin by reviewing the Stokes parameter description of polarization (\S\ref{sec:stokes})
and its relation
to $E$ and $B$ harmonic representation (\S\ref{sec:EB}).   We continue with a discussion of polarized Thomson scattering
in \S\ref{sec:thompol}.   In \S\ref{sec:acousticpol} and \ref{sec:gravwavepol} we discuss
the polarization signatures of acoustic oscillations and gravitational waves.

\subsection{Statistical Description}
\label{sec:stokes}

The polarization field can be analyzed in a way
very similar to the temperature field, save for one complication.
In addition to its strength, polarization also has an orientation,
depending on relative strength of two linear polarization states. 

The polarization field is defined locally in terms of Stokes parameters.
{In general, the polarization state of radiation in direction $\bn$ described by the
intensity matrix $\left< E_i(\bn) E_j^*(\bn) \right>$, where  ${\bf E}$ is the electric field
vector in the transverse plane and the brackets denote time averaging.}
{As a $2\times2$ hermitian matrix, it
can be decomposed into the Pauli basis}
\begin{eqnarray}
{\bf P} &=& C \left< {\bf E}(\bn) \, {\bf E}^\dagger(\bn) \right>   \nonumber\\
        &=& \Theta(\bn) {\sig{0}}
        + Q(\bn)  \, \sig{3}
        + U(\bn) \,  \sig{1}
        + V(\bn) \,  \sig{2}
\,, 
\end{eqnarray}
{where}
\begin{eqnarray}
\sig{0} & =& \mat{1}{0}{0}{1}   \,, \quad
\sig{2} = \mat{0}{-1}{1}{0} i  \,, \nonumber\\
\sig{1} &= &\mat{0}{1}{1}{0}  \,, \quad
\sig{3} = \mat{1}{0}{0}{-1} \,.
\end{eqnarray}
{Orthogonality of the Pauli matrices says that the Stokes parameters are 
recovered from the polarization matrix as ${\rm Tr}(\sigma_i {\bf P})/2$.} 
The Stokes $Q$ and $U$ parameter define the linear polarization state
whereas $V$ defines the circular polarization state.    We have chosen the
proportionality constant so that all the Stokes parameters are in
temperature fluctuation units.   

From this description, we see that $Q$ represents polarization aligned with one of the
principal axes of the transverse coordinate system whereas $U$ represents polarization at
$45^{\circ}$ to these axes.

\subsection{$EB$ Harmonic Description}
\label{sec:EB}

The disadvantage of the Stokes $Q$ and $U$ representation is that
the distinction between the two depends on the coordinate system for the
two transverse directions on the sky.    Under a rotation of this basis by $\theta$
\myequation{
Q'\pm i U' = e^{\mp 2i\theta} [Q \pm i U]\,.}

For a harmonic decomposition, it is more
useful to choose this basis to be given by the wavevector itself.
For small sections of the sky, the harmonic decomposition becomes a Fourier transform
and we can define the $E$ and $B$ harmonics as
\begin{eqnarray}
E(\bl) \pm i B(\bl) & = &   \int d \bn [Q'(\bn) \pm i U'(\bn)] e^{-i\bl \cdot \bn} \\
                            & = &  e^{\mp 2 i \phi_\bl} \int d \bn [Q(\bn) \pm i U(\bn)]   e^{-i\bl \cdot \bn} \,,\nonumber
\end{eqnarray}
where in the second line we have rotated $Q$ and $U$ back to a fixed coordinate
system with the 
angle $\phi_\bl$ that the Fourier vector makes with the $\vc{x}$ axis.

For linear scalar fluctuations, we have seen that the only direction for a given harmonic mode
is set by the direction of the wavevector itself: velocity fields point in this direction or its opposite,
quadrupole moments are symmetric about this axis, etc.  This means that symmetry 
requires that such sources only generate $Q'$  or $E$ for each mode.   
This symmetry also holds once all the modes are superimposed back to the full
polarization field:  linear scalar perturbations generate only $E$-modes where the polarization
direction is related to the direction in which the polarization amplitude changes.

To generalize this decomposition to the full curved sky, we need to replace plane waves,
the tensor eigenfunctions of the Laplace operator
in a flat space, to the correct tensor eigenfunctions for the curved space.
These are called the spin-2 spherical harmonics:
\myequation{ \nabla^2 {}_{\pm 2} Y_{\ell m} [\sig{3} \mp i \sig{1}] = -[l(l+1)-4] {}_{\pm 2}Y_{\ell m} 
[\sig{3} \mp i \sig{1}]\,.}
They obey the usual orthogonality and completeness relations
\begin{eqnarray}
\int d \bn {}_s Y_{\ell m}^*(\bn) {}_s Y_{\ell m}(\bn) &=& \delta_{\ell \ell'} \delta_{m m'}\,, \nonumber\\
\sum_{\ell m} {}_s Y_{\ell m}^*(\bn) {}_s Y_{\ell m}(\bn') &=& \delta(\phi - \phi') \delta(\cos\theta - \cos\theta') \,.
\end{eqnarray}
We can therefore decompose the linear polarization field just like the temperature field
\begin{eqnarray}
[Q(\bn) \pm i U(\bn)] = - \sum_{\ell m} [E_{\ell m} \pm i B_{\ell m}] {}_{\pm 2 } Y_{\ell m}(\bn)\,.
\end{eqnarray}
Likewise the power spectra are given by
\begin{eqnarray}
\left< E_{\ell m}^* E_{\ell m} \right> &= \delta_{\ell \ell'} \delta_{m m'} C_\ell^{EE}\,, \\
\left< B_{\ell m}^* B_{\ell m} \right> &= \delta_{\ell \ell'} \delta_{m m'} C_\ell^{BB} \,,
\end{eqnarray}
and the cross correlation of $E$-polarization with temperature by
\myequation{
\left< \Theta_{\ell m}^* E_{\ell m} \right> = \delta_{\ell \ell'} \delta_{m m'} C_\ell^{\Theta E}\,.}
{Other cross correlations  vanish if parity is conserved.}

\begin{figure}[tb]
\begin{center}
\includegraphics[width=2in]{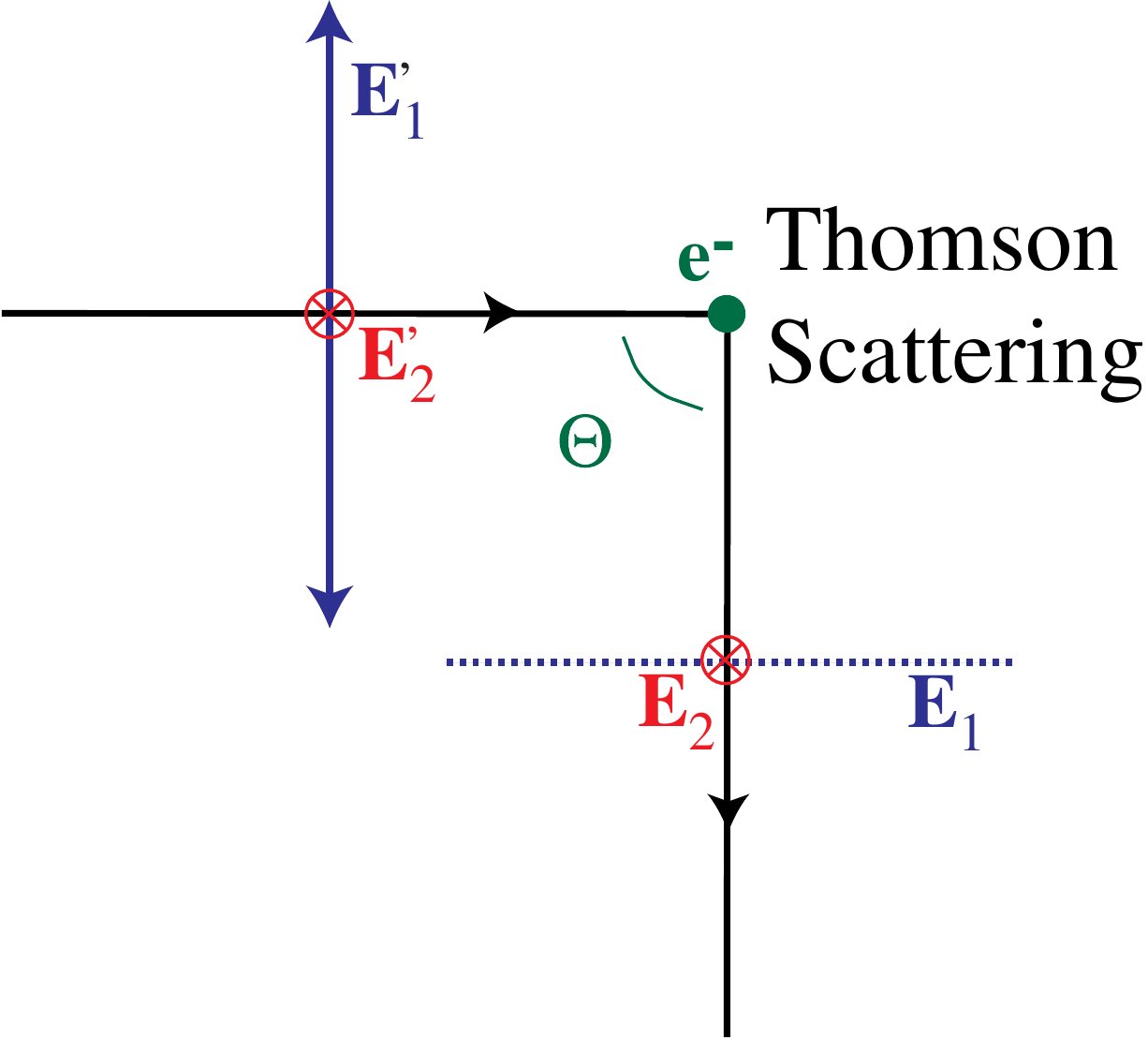}
\caption{Thomson scattering geometry.  A quadrupole anisotropy in the incoming radiation
leads to linear polarization.  For scattering at $\Theta=\pi/2$, only one component of the
initially unpolarized radiation (${\bf E}_2'$) is scattered leaving one outgoing state
(${\bf E}_1$) unpopulated.} \label{fig:thomsonpolarization}
\end{center}
\end{figure}

\subsection{Thomson Scattering}
\label{sec:thompol}

The differential cross section for Thomson scattering
\myequation{
{d \sigma \over d\Omega} = {3 \over 8\pi} |\hat {\bf E}' \cdot \hat{\bf E}|^2\sigma_T
\,,
\label{eqn:Thomson}
}
 is polarization dependent and
hence scattering generates polarization.
Here
$\hat {\bf E}'$ and $\hat {\bf E}$ denote the incoming and outgoing
directions of the electric field or polarization vector.
Consider incoming radiation in the $\hat\vc{x}$ direction scattered at right
angles into the $-\hat\vc{y}$ direction (see Fig.~\ref{fig:thomsonpolarization}).
Heuristically, incoming radiation shakes an electron in the direction
of its electric field vector or polarization $\hat\vc{e}'$
causing it to radiate with an outgoing polarization parallel to that
direction.  However since the outgoing polarization 
$\hat\vc{e}$
must be orthogonal to the outgoing direction, incoming radiation that
is polarized parallel to the outgoing direction cannot scatter leaving 
only one polarization state.

The incoming radiation however comes
from all angles.  If it were completely isotropic in intensity, 
radiation coming along the $\hat \vc{z}$ would provide the polarization
state that is missing from that coming along $\hat\vc{x}$ 
leaving the net outgoing radiation unpolarized.
Only a quadrupole temperature anisotropy in the radiation
generates a net linear polarization from Thomson scattering.
As we have seen, a quadrupole 
can only be efficiently generated
if the universe is optically thin to Thomson 
scattering a given perturbation.

\subsection{Acoustic Polarization}
\label{sec:acousticpol}

Acoustic oscillations in the dissipation regime provide the conditions
necessary for polarization.
Recall that radiative viscosity in the plasma is equivalent to quadrupole
anisotropy in the photons.  
Since the quadrupole is of order (see Fig.~\ref{fig:osc3})
\begin{equation}
\pi_{\gamma} \sim {kv_\gamma \over \dot\tau} \sim \left({k \over k_{D}}\right)
 {v_\gamma \over k_D \eta_*}  \,,
\end{equation} 
the polarization
spectrum rises as $l/l_D$ to peak at the damping scale with an amplitude of
about 10\% of the temperature fluctuations before falling due to the elimination of
the acoustic source itself due to damping.
Since $v_\gamma$ is out of
phase with the temperature, 
\myequation{\Theta+\Psi \propto \cos(ks); \quad v_\gamma \propto \sin(ks)\,, }
the polarization peaks are also out of phase with the
temperature peaks. 
Furthermore, the phase relation also tells us that the polarization is 
correlated with the temperature perturbations.  The correlation power $C_l^{\Theta E}$ 
being the product of the two, exhibits oscillations at twice the acoustic frequency
\myequation{(\Theta+\Psi)(v_\gamma) \propto \cos(ks)\sin(ks) \propto \sin(2 ks)\,.}
As in the case of the damping, the predicting the precise value requires
numerical codes (\citealt{BonEfs87})
since $\dot\tau$ changes so rapidly near recombination.  
Nonetheless the detailed predictions bear these
qualitative features.

\begin{figure}[tb]
\begin{center}
\includegraphics[width=4.5in]{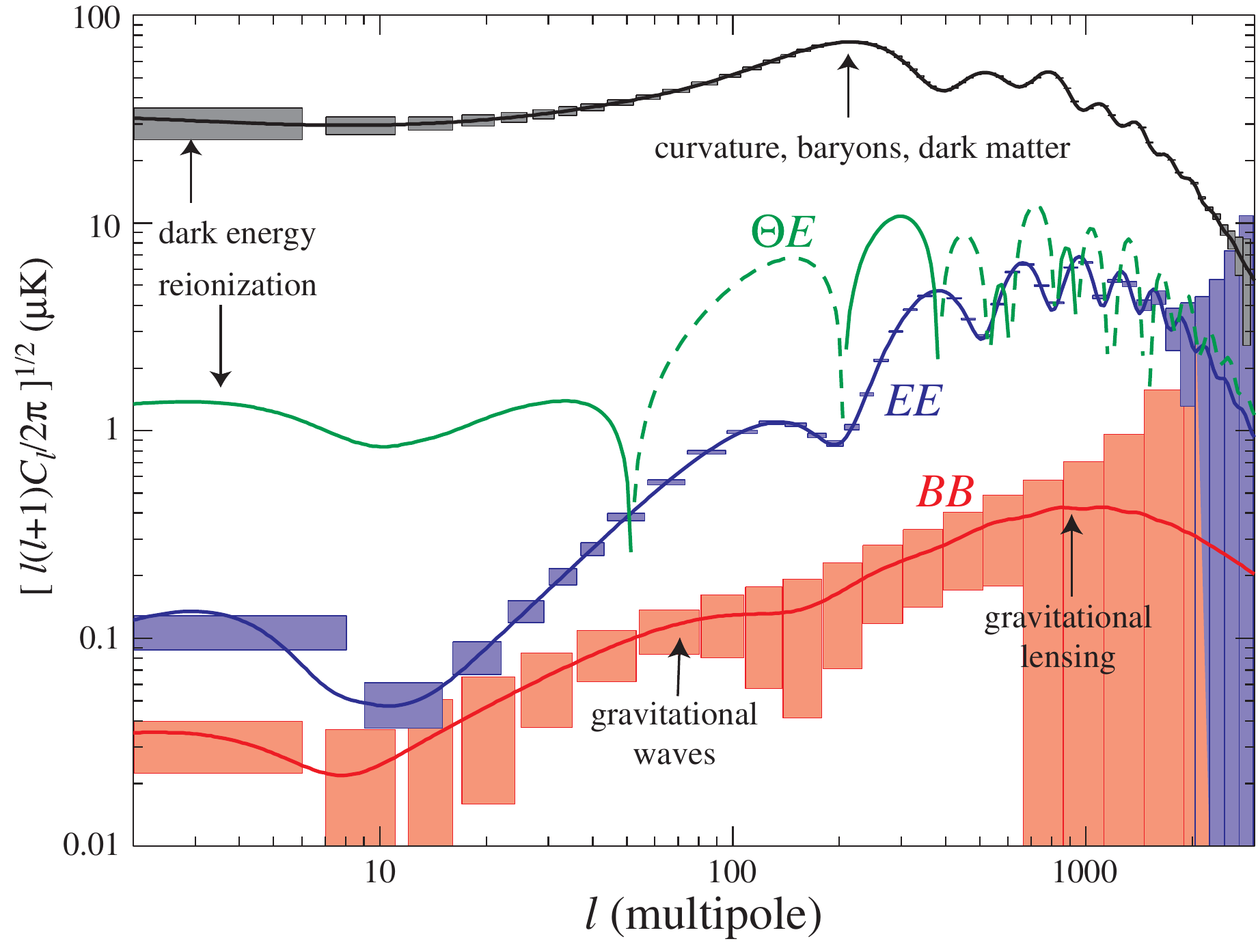}
\caption{Polarized landscape.  While the $E$-spectrum and $\Theta E$ cross correlation are
increasingly well measured, the $B$-spectrum from inflationary gravitational waves (shown here
near the maximal value allowed by the temperature spectrum) and
gravitational lensing remains undetected.  Shown here are projected error bars associated with Planck sample
variance and detector noise.  Adapted from \cite{HuDod02}.} \label{fig:spectra}
\end{center}
\end{figure}

 Like the 
damping scale, the acoustic polarization spectrum is uniquely predicted 
from the temperature spectrum once
$\Omega_b h^2$, $\Omega_m h^2$ 
and the initial conditions are specified.   Polarization thus
represents a sharp test on the assumptions of 
the recombination physics and power law curvature fluctuations in 
the initial conditions used in interpreting the temperature peaks.  
For example, features in the initial power spectrum appear more distinct
in the polarization than the temperature due to projection effects
(see \emph{e.g.} \cite{HuOka03}).

\subsection{Gravitational Waves}
\label{sec:gravwavepol}

During the break down of tight coupling that occurs at last scattering,
any gravitational waves present will also imprint a local quadrupole anisotropy 
to the photons and hence a linear polarization to the CMB (\citealt{Pol85}).  
These contribute to
the $BB$ power and their detection would provide invaluable information on the
origin of the fluctuations.  
Specifically, in simple inflationary models their
amplitude gives the energy scale of inflation.  The gravitational wave amplitude $h$
oscillates and decays once inside the horizon, so the associated polarization source
scales as $\dot h /\dot \tau$ and so peaks at the $l \approx 100$
horizon scale and not the damping scale at recombination (see Fig.~\ref{fig:spectra}).
This provides a useful scale separation of the various polarization effects.

If the energy scale of inflation is near the $10^{16}$GeV scale then the signal
is potentially detectable by the next generation of polarization experiments.

\section{Discussion}
\label{sec:discussion}

The one and a half decades since the discovery of CMB anisotropy by COBE DMR has
seen remarkable progress that ushered in the current epoch of precision cosmology.
From the preliminary detections of degree scale power to the current measurements
of 5 acoustic peaks, the damping tail and acoustic polarization, the milestones in
the observational verification of our theoretical understanding of the universe at recombination
have steadily been overtaken.  Correspondingly, the measurements of the energy
density contents of the universe at recombination and the distance to and hence
expansion rate since recombination have improved from order unity constraints to 
several percent level determinations.

The next generation of experiments, including the Planck satellite and ground based
polarization measurements, will push these determinations to the 1\% level
and beyond and perhaps detect the one outstanding prediction of the recombination
epoch: the $B$-mode polarization of gravitational waves from inflation.   Likewise the
enhanced range of precision measurements of the power spectrum will bring 
measurements of the spectrum of scalar perturbations to the percent level and
further test the physics of inflation.

Beyond the recombination epoch reviewed here, these experiments will also test secondary
temperature and polarization anisotropy from reionization, lensing, and galaxy clusters
as well as the Gaussianity of the initial conditions.    Combined, these measurements will test
the standard cosmological model with unprecedented precision.

\medskip
\noindent{\emph{Acknowledgments:} I thank the organizers, R. Rebelo and J.A. Rubino-Martin, as well as
the students of the XIX Canary Island Winter School of Astrophysics.}

\bibliography{canaries_wh}
\bibliographystyle{canaries_wh}

\end{document}